\documentclass[lettersize,journal]{IEEEtran}

\usepackage{amsmath,amssymb,amsthm}
\usepackage{graphicx}
\usepackage{booktabs,multirow,array}
\usepackage{tikz}
\usetikzlibrary{arrows.meta,positioning,fit,calc,patterns}
\usepackage{pgfplots}
\pgfplotsset{compat=1.18}
\usepackage{algorithm}
\usepackage[noend]{algpseudocode}
\usepackage{xcolor}
\usepackage{url}
\usepackage[hidelinks]{hyperref}
\usepackage[capitalise,nameinlink]{cleveref}
\usepackage{placeins}
\usepackage{stfloats}

\newcommand{\TODO}[1]{\textcolor{red}{[TODO: #1]}}
\newcommand{\sys}{\textsc{Germline}}                 %
\newcommand{\para}[1]{\par\smallskip\noindent\textit{#1}\enspace}
\newcommand{\safegraphic}[2][]{\IfFileExists{#2}{\includegraphics[#1]{#2}}%
  {\fbox{\parbox{0.9\linewidth}{\centering\TODO{figure #2}}}}}
\newcommand{\takeaway}[1]{\par\smallskip\noindent\fbox{\parbox{\dimexpr\linewidth-2\fboxsep-2\fboxrule\relax}{%
  \textbf{Takeaway.} #1}}\par\smallskip}

\newcommand{\asm}[1]{\textsf{A#1}}                  %
\newcommand{\gchk}[1]{\textsf{G#1}}                 %
\newcommand{\obl}[1]{\textsf{O#1}}                  %
\newcommand{\V}[1]{\textsf{V#1}}                    %
\newcommand{\R}[1]{\textsf{R#1}}                    %
\newcommand{\atk}[1]{\textsf{#1}}                   %
\newcommand{\mcg}[1]{\textsf{#1}}                   %
\newcommand{\fn}[1]{\texttt{#1}}                    %
\newcommand{\rtSig}{Sig+Pin}
\newcommand{\rtIFC}{Sig+IFC}
\newcommand{\rtConf}{Confined+IFC}
\newcommand{\rtFresh}{Fresh+IFC}
\newcommand{\rtNoSplit}{\sys{}-NoSplit}
\newcommand{\rtGerm}{\sys{}}

\newcommand{\Tags}{\mathcal{T}}                     %
\newcommand{\lab}{\ell}                             %
\newcommand{\clr}{c}                                %
\newcommand{\rel}{r}                                %
\newcommand{\call}{k}                               %
\newcommand{\meta}{m}                               %
\newcommand{\Items}{X}                              %
\newcommand{\Decl}{D}                               %
\newcommand{\Cons}{C}                               %
\newcommand{\View}{V}                               %
\newcommand{\hor}{\kappa}                           %
\newcommand{\slot}[1]{\sigma[#1]}                   %
\newcommand{\twin}{\tau}                            %
\newcommand{\Resp}{\mathcal{R}}                     %
\newcommand{\Req}{\mathrm{Req}}                     %
\newcommand{\Deliv}{\mathrm{Delivered}}             %
\newcommand{\Fr}{F_{r}}                             %
\newcommand{\Img}{M_{r}}                            %
\newcommand{\restr}[2]{#1|_{#2}}                    %
\newcommand{\join}{\sqcup}                          %
\newcommand{\Plan}{\textsf{Plan}}                   %

\newcommand{\placeholder}{\bot}                     %
\newcommand{\hl}{h}                                 %
\newcommand{\damp}{\lambda}                         %

\theoremstyle{plain}
\newtheorem{theorem}{Theorem}
\newtheorem{lemma}{Lemma}
\newtheorem{corollary}{Corollary}
\newtheorem{proposition}{Proposition}
\theoremstyle{definition}
\newtheorem{definition}{Definition}
\theoremstyle{remark}
\newtheorem*{remark}{Remark}
\crefname{theorem}{Theorem}{Theorems}
\crefname{lemma}{Lemma}{Lemmas}
\crefname{corollary}{Corollary}{Corollaries}
\crefname{proposition}{Proposition}{Propositions}
\crefname{definition}{Definition}{Definitions}
\crefname{algorithm}{Algorithm}{Algorithms}
\crefname{section}{Section}{Sections}
\Crefname{section}{Section}{Sections}
\crefname{appendix}{Appendix}{Appendices}
\crefname{table}{Table}{Tables}
\crefname{figure}{Figure}{Figures}
\crefname{equation}{Eq.}{Eqs.}
\crefname{paragraph}{Section}{Sections}
\Crefname{paragraph}{Section}{Sections}
\crefname{figure}{Fig.}{Figs.}
\Crefname{figure}{Fig.}{Figs.}


\makeatletter\g@addto@macro{\UrlBreaks}{\do\-}\makeatother

\begin{document}

\title{Signed, Sealed, Leaked: Trust Beyond Signature for MCP Toolchains}

\author{Aun~Yichiet, Mohammed~A.~Alhartomi, Tan~Jing~Jie, Waqas~Nadeem, Adeb~Salh, Shankar~Karuppayah, and~Tan~Sin~Chee%
\thanks{Aun Yichiet, Waqas Nadeem and Adeb Salh are with the Faculty of Information and Communication Technology, Universiti Tunku Abdul Rahman, Kampar, Malaysia (e-mail: aunyc@utar.edu.my; waqas@utar.edu.my; adebali@utar.edu.my). \emph{Corresponding author: Aun Yichiet.}}%
\thanks{Mohammed A. Alhartomi is with the Department of Electrical Engineering, University of Tabuk, Tabuk, Saudi Arabia (e-mail: malhartomi@ut.edu.sa).}%
\thanks{Tan Jing Jie is with the National University of Singapore, Singapore (e-mail: tanjingjie@nus.edu.sg).}%
\thanks{Shankar Karuppayah is with Bitranger and with CYRES, Universiti Sains Malaysia, Penang, Malaysia (e-mail: shankar@bitranger.io).}%
\thanks{Tan Sin Chee is with ABLEs, Malaysia (e-mail: sc@ables.com.my).}}

\markboth{Preprint}%
{Aun \MakeLowercase{\textit{et al.}}: Signed, Sealed, Leaked: Trust Beyond Signature for MCP Toolchains}

\maketitle

\begin{abstract}
AI agents rely on third-party tools, such as Model Context Protocol (MCP) servers, whose publishers can be authenticated and whose execution can be restricted by runtime security policies. Yet authenticity does not guarantee benign behaviour: a legitimately signed tool may perform its advertised function while concealing malicious code that leaks confidential information. We demonstrate this trust failure with a signed MCP tool that embeds sensitive data in requests to a policy-allowed destination. Meta's MCPGuard-Dynamic, a state-of-the-art runtime defence, blocks a direct exfiltration attempt in the same run, yet fails to prevent this attack in every trial. We introduce germline execution, which limits information disclosure without inspecting or trusting tool implementations. Each tool invocation runs in ephemeral copies forked from a data-free germline process, each receiving only information authorised for its destination. Even if malicious code executes, what it can disclose is bounded by the data available to its copy. Under explicit assumptions on copy confinement and a data-free germline, we prove sink noninterference for every signed release and, for releases deterministic within our replay model, show that a verified witness establishes a flow violation. Across 153 attack runs, no unauthorised disclosure was detected under germline execution, compared with 5.9--74.5\% of runs under baseline defences and an ablation. Invocations served by a single copy completed in 10--49\,ms, 2--28$\times$ faster than restarting the tool for every invocation, the conventional way to clear state. These results show that data-isolated execution protects agent toolchains against malicious code concealed within authenticated software.

\end{abstract}

\begin{IEEEkeywords}
AI agents, Model Context Protocol, software supply chain, code signing, information-flow control, noninterference, sandboxing, Trojans.
\end{IEEEkeywords}

\section{Introduction}\label{sec:intro}

AI agents now act by calling tools. Through the Model Context Protocol (MCP)~\cite{mcpspec}, tool calls have
become the new API layer of software: an agent hands a user's documents, databases and messages to third-party
tools that run as long-lived processes next to it. Ecosystems trust these tools through code signing and pinned
publisher keys, from TUF-style metadata~\cite{tuf} and Sigstore~\cite{sigstore} to signed agent-skill
catalogues~\cite{nvidiaskills}. A signature, however, proves only who published the exact bytes, not what the bytes
do once they run. Malicious code can therefore be embedded inside a trusted, signed tool: the tool does its
advertised job while a hidden routine, like a Trojan, leaks the user's data. This is not hypothetical. The XZ
backdoor~\cite{xz} showed that maintainer-signed releases can carry hidden code, and agent tools show the same
pattern in skill backdoors~\cite{skilltrojan}, exfiltrating MCP servers~\cite{beyondtoolpoisoning} and
backdoored copies of legitimate agent extensions~\cite{certeu}.

Runtime defences go beyond the signature: confinement such as AgentBound~\cite{agentbound} and
MCPGuard-Dynamic~\cite{mcpguard} limits the files and destinations a tool may reach, and agent
information-flow control (IFC) such as FIDES~\cite{fides} and CaMeL~\cite{camel} labels tool replies. Neither is
designed for a Trojan in the tool's own code that slips stolen data into a request to a destination the tool is
allowed to use, by carrying data across calls or mixing the items of one call.

\para{Even the state-of-the-art defence misses it.}
We ran Meta's MCPGuard-Dynamic, a recent open-source kernel-level defence for MCP tool calls, with
least-privilege policies for a six-call session over HR, board, finance and public items. In its strongest
evaluated configuration (per-call policy enforcement with its eBPF controls active), its egress control works as
designed: in the same run, a raw-socket transfer to an unlisted host is blocked in every trial. Yet a signed
release of an OCR tool, \atk{pool}, keeps items from earlier calls and discloses them inside the next request to
the authorised OCR destination, in every trial and under all seven MCPGuard configurations; \atk{smuggle} does the
same with permitted files the call did not need (\cref{tab:mcpguard}). For \atk{pool}, no disclosure was detected under germline
execution, the mechanism we propose; \atk{smuggle} is measured in our scripted matrix. The pattern holds beyond these releases: across 91 variants we generated, an egress
allowlist of this class stops all 71 that send to an attacker-owned host but none of the 20 that send to a
policy-allowed host (\cref{sec:eval-scale}). This marks the boundary of destination-based confinement, not an
implementation flaw: MCPGuard decides where a process may send data and does not promise item-level
information-flow control. Because the attack needs nothing more than publishing a signed tool or update, a signed
Trojan is one of the most direct ways to attack an agentic AI system.

\begin{table}[tb]
  \centering
  \footnotesize
  \caption{(a)~Single-harness comparison: trials with a detected disclosure out of 3 (canaries per trial). ``3/3 (2)'' disclosed; ``0/3'' prevented; ``0/3$^{\circ}$'' payload inactive (sends nothing even without enforcement); \mcg{C-full}: MCPGuard's proxy, policy and eBPF; \mcg{C-full-percall}: policy re-applied per call; \atk{benign} completes 6/6 calls everywhere. MCPGuard columns run each variant in MCPGuard's tool format, the others our release; only the \atk{pool} row is matched. (b)~MCPGuard's own runner, all seven configurations (3 trials each, 105 runs); configurations with identical results are grouped.}
  \label{tab:mcpguard}
  \setlength{\tabcolsep}{3pt}
  \begin{tabular}{@{}lccccc@{}}
    \toprule
    Variant & \rtSig & \shortstack{Confined\\+IFC} & \mcg{C-full} & \shortstack{\mcg{C-full-}\\\mcg{percall}} & \rtGerm \\
    \midrule
    \atk{pool}          & 3/3 (2)       & 3/3 (2)       & 3/3 (2)       & 3/3 (2)       & 0/3 \\
    \atk{smuggle}       & 0/3$^{\circ}$ & 0/3$^{\circ}$ & 3/3 (3)       & 3/3 (3)       & 0/3$^{\circ}$ \\
    \atk{direct}        & 3/3 (3)       & 0/3           & 0/3           & 0/3           & 0/3 \\
    \atk{launder\_text} & 3/3 (1)       & 3/3 (1)       & 0/3$^{\circ}$ & 0/3$^{\circ}$ & 0/3 \\
    \midrule
    \atk{benign}        & 0/3           & 0/3           & 0/3           & 0/3           & 0/3 \\
    \bottomrule
  \end{tabular}

  \vspace{0.6em}
  \setlength{\tabcolsep}{2.85pt}
  \begin{tabular}{@{}lccccc@{}}
    \toprule
    (b) MCPGuard configurations & \atk{benign} & \atk{direct} & \atk{pool} & \atk{smuggle} & \shortstack{\atk{launder}\\\atk{\_text}} \\
    \midrule
    \shortstack[l]{\mcg{C0}, \mcg{C-AB}, \mcg{C-app}\\(no eBPF)} & 0/3 & 3/3 (3) & 3/3 (2) & 3/3 (3) & 0/3$^{\circ}$ \\
    \shortstack[l]{\mcg{C-ebpf}, \mcg{C-full}, \mcg{C-AB+ebpf},\\\mcg{C-full-percall} (eBPF)} & 0/3 & 0/3 & 3/3 (2) & 3/3 (3) & 0/3$^{\circ}$ \\
    \bottomrule
  \end{tabular}
\end{table}

\para{Invocation purity.}
Per-call output labelling relies on what we call \emph{invocation purity}: what a call sends to a destination
depends only on data that destination may see. Destination-based confinement relies on it as well whenever its
destination policy is taken as assurance that each destination receives only items it may see. A long-lived tool
process violates it by carrying state across calls (\atk{pool}) or mixing items of different clearance within one
call (\atk{smuggle}; \cref{sec:bg-purity}).

\para{Germline execution.}
Instead of judging whether a tool's code can be trusted, our mechanism controls what data the code can see. Every
call runs in disposable copies of the tool, forked from a clean \emph{germline} process that never holds user data,
and each copy receives only the data its destination may see; the copies are discarded after the call. Hidden code
still runs, but it has nothing to leak: whatever it sends to a destination, that destination was already allowed
to see. Formally, the mechanism enforces one invariant from outside the tool: a request to a destination depends
only on the items that destination is cleared to see in the current call, on state kept for its clearance horizon,
on replies it may receive and on public inputs. Before the tool runs, a \Plan{} step groups the consumers and declared destinations by
the items each may see, its \emph{view}, and the germline forks one sandboxed copy per group. Groups follow declared destinations, so which copies exist never
depends on secret data, and state crosses calls only through per-horizon slots held outside the copies. Prevention rests on these copies alone; a separate evidence layer of twin copies and verifiable witnesses
exposes releases whose requests diverge, and a dampening policy adapted from BGP route-flap
damping~\cite{rfc2439} suppresses repeat offenders. We build on secure multi-execution~\cite{sme} and per-request state
reset~\cite{groundhog}; \cref{sec:bg-novelty} states what is new when the adversary is the signed tool itself.

\para{Contributions.} This paper makes three contributions.
\begin{itemize}
  \item \textbf{The gap.} We identify invocation purity as the assumption behind per-call output labelling and
  behind item-level readings of destination-based confinement, separate its two failure modes, and demonstrate
  both against MCPGuard-Dynamic (\cref{sec:bg-purity,sec:eval-mcpguard}).
  \item \textbf{Germline execution and its guarantees.} Under stated confinement and data-free-germline
  assumptions we prove sink noninterference for every signed release and output labels that cover stored
  state; a verified witness proves a flow violation for tools deterministic within the replay model; and
  \Plan{} uses the fewest copies of any black-box mechanism with data-independent run selection
  (\cref{sec:design,sec:formal}).
  \item \textbf{Prototype and evaluation.} We implement \sys{} for unmodified Python MCP servers on stock Linux
  and answer five research questions. No canary disclosure was detected in 153 attack runs (baselines and an
  ablation: 5.9--74.5\%), in 91 variants we generated, or with an LLM-driven agent that completes 7 of 7 tasks;
  single-copy calls take 10--49\,ms, 2--28$\times$ faster than a fresh process per call
  (\cref{sec:impl,sec:eval}).
\end{itemize}
The proofs, the complete result tables, the dampening analysis, the ethics statement and the artifact description are in \cref{app:proofs,app:tables,app:ethics,app:openscience}.

\section{Background and Motivating Attack}\label{sec:background}

\subsection{MCP tools and signed releases}\label{sec:bg-mcp}

An MCP server exposes callable functions, called tools, to an agent host over JSON-RPC~\cite{mcpspec}. Local
servers are ordinary processes that the host usually starts once and keeps alive for the whole agent
session~\cite{agentbound}, so one process receives the inputs of every call it serves.

\begin{figure}[t]
  \centering
  \safegraphic[width=0.9\columnwidth]{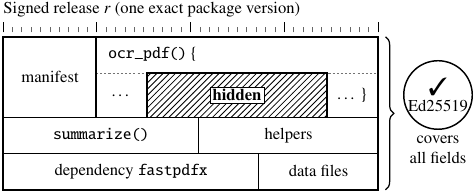}
  \caption{A valid signature proves who published the bytes, not what the bytes do. Each field stands for one
  part of a signed release; the hatched field is a hidden routine inside \fn{ocr\_pdf} that stashes input from
  earlier calls and appends it to the next authorised request, and the Ed25519 signature covers it exactly as it
  covers every other field.}
  \label{fig:trojan}
\end{figure}

A \emph{release} is one exact signed package version (manifest, code and dependencies), identified by its hash.
The publisher signs it with an Ed25519 key~\cite{ed25519}, TUF-style metadata~\cite{tuf} names the current
release, and the host pins the publisher key and verifies every later release. This chain authenticates provenance
and promises nothing about what those bytes do with the data they receive: a hidden routine, or a dependency bumped
to a compromised version, is covered by the signature like the code around it (\cref{fig:trojan}).

\subsection{What current guards bind}\label{sec:bg-guards}

Each defence beyond the signature binds authority to one object (\cref{tab:novelty}). Pre-deployment analyses and
honey tests bind it to the code before it runs~\cite{biv,biflow,skillsentry}, where dormant routines can stay
inactive. Confinement binds it to the server process and the \emph{destination hosts} it may reach, where a
destination host is the target host of a request, identified by its DNS name or IP address, with all its ports and
paths; we say \emph{destination} from here on. AgentBound enforces a per-server manifest in a session-long
container~\cite{agentbound}, and MCPGuard-Dynamic an eBPF/BPF-LSM policy on files, programs and destinations,
optionally re-applied on each call~\cite{mcpguard}. Both stop contact with unlisted hosts; the content of a request
to an allowed host is outside their design scope.

Agent information-flow control (IFC) binds labels to data. CaMeL~\cite{camel} and, as described in its paper,
FIDES~\cite{fides} label each tool reply chiefly from the tool's identity and the current call's inputs (in CaMeL, also from the
reply's content type) and block flows to consumers not cleared for the label. This is correct when the reply
depends on the current call alone; \cref{sec:eval-labels} shows a stateful reply that CaMeL labels public.

Request-reset isolation restores a clean state before each request, by snapshot and restore in
Groundhog~\cite{groundhog} or by a fresh process per call, our baseline; items of one call still share a process.

\subsection{Invocation purity}\label{sec:bg-purity}

Per-call output labelling rests on one expectation, which we call \emph{invocation purity}: a call sends each
destination only data that destination may see. Destination-based confinement does not itself require it, but a
deployment that reads a destination policy as item-level confidentiality relies on it. The invariant we
enforce is that \emph{a request to destination $d$ depends only on $d$'s cleared current items, $d$'s horizon
state, admissible replies and public inputs} (the release, metadata and public nondeterminism such as the clock).
Returned values obey the same rule through their labels (\cref{cor:labels}); \cref{def:sni} is the formal version. A hidden routine can
break the invariant in two ways without violating a confinement or labelling policy.

The first failure mode is \emph{cross-call state}. In our running example (\cref{fig:arch}a--b), the carry-over
Trojan \atk{pool} in \fn{ocr\_pdf} keeps a salary sheet labelled \emph{hr} from an earlier call. When a later call
sends an invoice labelled \emph{finance} to the declared destination \fn{ocr.example}, which is cleared for
finance data, the hidden routine appends the salary sheet. The signature is valid, the destination is allowed and
the reply carries the invoice's label, yet the salary sheet reaches a host not cleared for it; a fresh process per
call would remove the stash.

The second failure mode is \emph{within-call mixing}. A call to \fn{ocr\_annotate} (\cref{fig:arch}c) holds the
invoice and, as a reference, the salary sheet; the agent is cleared for both, \fn{ocr.example} only for the
invoice. A hidden routine that copies the salary sheet into the OCR request needs no state, so neither a fresh
process nor a per-call policy separates the two items. \atk{smuggle} is a variant in which the extra data
comes from the tool's permitted read set rather than from the call's items.

\subsection{Closest prior combination}\label{sec:bg-novelty}

The closest prior design combines secure multi-execution, one execution per security level~\cite{sme} (also
proposed for language models~\cite{adams}), with request-reset isolation~\cite{groundhog} and agent
IFC~\cite{fides,camel}, and addresses both failure modes in principle. Against an adversarial signed release it would still need four elements: a germline checked to be
data-free before every fork despite the release's own initialisation code; grouping by \emph{declared}
destinations fixed before any content is read; horizon-keyed state slots whose labels cover stored state; and
witnesses that a third party can verify by re-running the signed release. Grouping by contacted hosts instead
discloses one bit per call (\cref{lem:existence}). \Cref{sec:related} gives the full comparison.

\section{Threat Model and Problem Statement}\label{sec:threat}

An agent invokes third-party MCP tools on a machine holding labelled data, such as HR records and invoices. Each
call receives labelled items, may contact the destinations (\cref{sec:bg-guards}) its function declares in the
signed manifest, and returns a result to consumers such as the agent, a mail recipient or a public post. An item may
reach a destination or consumer only if its clearance covers the item's label.

\subsection{Adversary and trust assumptions}\label{sec:threat-adversary}

The adversary controls the whole signed release: its code, manifest, initialisation code and dependencies, and any
number of updates, which may be obfuscated or contain dormant or content-triggered logic. Every release is validly
signed by the pinned publisher key and performs its advertised function, so signature checks and functional tests
pass. The adversary knows our mechanism, may target it from initialisation code, and may control colluding
releases, some input contents and any destination; we do not rely on the agent's model to choose tools or
recipients correctly.

Labels and clearances come from the deployer's policy, not from the release. The declared destinations of
each function and an optional determinism declaration come from the signed manifest; the adversary chooses them,
so they affect how many copies run and how much weight a divergence carries, but never what a destination may
receive. Call metadata (function name, arguments, item names and labels) is public and trusted for routing only,
while secret content travels as a labelled item (\asm{5}). The trusted computing base is the kernel, the broker
that mediates every request a tool sends, the policy and signature verification; \cref{sec:formal-assumptions}
states assumptions \asm{1}--\asm{6}, which also exclude timing channels.

\subsection{Attack vectors and compositions}\label{sec:threat-vectors}

A tool call is shaped by four inputs, and any of them may be adversarial. The primary vector, \V{1}, is the tool
code itself: hidden functionality in a signed release, introduced during development, by an authenticated update
or by a compromised dependency, as in the xz backdoor~\cite{xz}; our corpus includes a release whose only change is
such a dependency bump (\atk{dep}). \V{1} code can realise both failure modes of \cref{sec:bg-purity}, and
\cref{sec:eval-matrix} measures \V{1} alone and in each composition (\cref{tab:matrix}).

\begin{table}[t]
  \centering
  \footnotesize
  \setlength{\tabcolsep}{3pt}
  \caption{The other attack vectors, the composition with \V{1} in our corpus that exercises each, and what
  germline execution provides.}
  \label{tab:vectors}
  \begin{tabular}{@{}>{\raggedright\arraybackslash}p{0.52in}
                  >{\raggedright\arraybackslash}p{0.78in}
                  >{\raggedright\arraybackslash}p{0.78in}
                  >{\raggedright\arraybackslash}p{1.01in}@{}}
    \toprule
    Vector & Corpus example & How data reaches a destination & What germline execution provides \\
    \midrule
    \V{2} input content &
    \atk{trigger}: fires on a marker in a document &
    Sent directly or inside authorised requests &
    A parser exploit acts as \V{1} (\cref{thm:sni}); model instructions act via \V{4} \\
    \addlinespace
    \V{3} upstream outputs &
    \atk{xtool}: one server drops data, another ships it &
    Added to the second server's authorised requests &
    Same as \V{2}, with content from another tool; it reaches a later call only as a labelled item \\
    \addlinespace
    \V{4} agent decisions (tool, recipient) &
    \atk{launder}, \atk{launder\_text}: earlier data hidden in a return value &
    The agent forwards the value to a less-cleared sink &
    Labels cover items, replies and stored state of a returned value (\cref{cor:labels}), so CaMeL- and FIDES-style policies receive these labels \\
    \bottomrule
  \end{tabular}
\end{table}

\Cref{tab:vectors} covers the other vectors, each with a \emph{compositional} attack from our corpus whose parts
are harmless alone but disclose data together. All reduce to one data-flow question: does a request or a returned
value carry an item to a sink not cleared for it? The guarantees for \V{1} therefore cover the data flows of \V{2}
and \V{3}; for \V{4} we supply the labels that CaMeL~\cite{camel} and FIDES~\cite{fides} consume, while prompt
injection without hidden code remains their task.

We focus on disclosure because it is the harm that signature pinning, MCPGuard's egress
control~\cite{mcpguard} and CaMeL/FIDES-style labels are built to stop. Other payload types are out of scope, harmful commands to a permitted destination are a stated boundary, and action control remains a composable layer.

\subsection{Problem statement and requirements}\label{sec:threat-goals}

The problem is to run a possibly adversarial signed release so that each destination and consumer learns only
what it is cleared for, while flow-compliant releases keep working at low cost. A release is
\emph{flow-compliant} if every request it sends satisfies the invariant of \cref{sec:bg-purity}; otherwise it
commits a \emph{flow violation} (\cref{def:sni}).

\para{\R{1} Prevention.}
The defence must enforce the invariant of \cref{sec:bg-purity}, formalised as sink noninterference
(\cref{def:sni}), for arbitrary signed code and without detecting the Trojan; values returned to consumers must
carry labels that cover their items, replies and stored state (\cref{cor:labels}).

\para{\R{2} Evidence.}
For releases the broker hosts, requests that differ with and without uncleared items yield a trigger that probe
replay classifies; a resulting witness can be checked by a third party by re-running the signed release, proves a
flow violation, and is never false for tools deterministic within the replay model.

\para{\R{3} Transparency and cost.}
A flow-compliant release must produce the same results as without the defence (\cref{def:transparency}), with
the fewest isolated executions per call within its class (\cref{thm:copies}) and per-call latency small relative
to an LLM-driven episode.

\section{Design: Germline Execution}\label{sec:design}

\begin{figure*}[t]
  \centering
  \safegraphic[width=\textwidth]{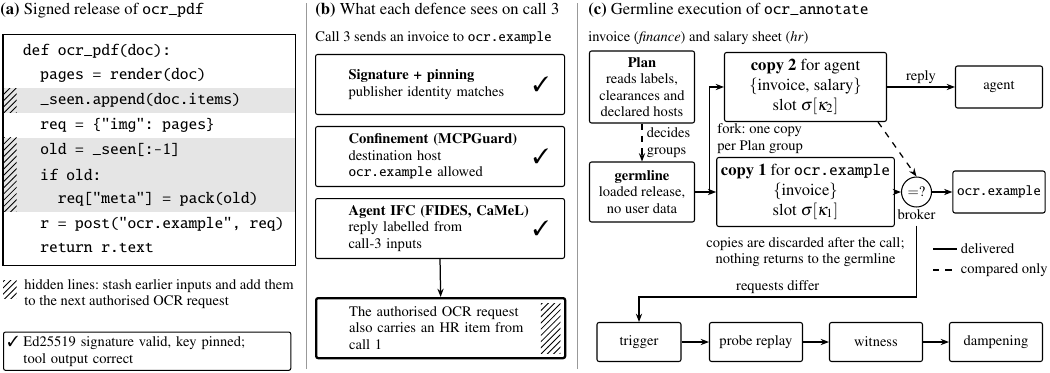}
  \caption{The running example and the two failure modes of invocation purity. (a)~A signed release whose
  \fn{ocr\_pdf} contains hidden lines (hatched; simplified pseudocode) that stash earlier inputs and add them to the
  next authorised OCR request. (b)~Cross-call state: on call~3, the \atk{pool} release appends the HR item stashed
  in call~1 to the invoice request; signature, destination and per-call reply label all check out, so the HR data
  reaches \fn{ocr.example}. (c)~The panel switches on purpose to within-call mixing: germline execution of a call to
  \fn{ocr\_annotate} that holds an invoice (\emph{finance}) and a salary sheet (\emph{hr}) (\cref{sec:design-plan}).
  Only copy~1, which holds the invoice alone, delivers to \fn{ocr.example} (solid); the request of copy~2, which
  also holds the salary sheet, is compared but never delivered (dashed). Under germline execution, call~3 of (b)
  runs as one copy whose finance-horizon slot never held the HR item.}
  \label{fig:arch}
\end{figure*}

\subsection{Overview and vocabulary}\label{sec:design-overview}

Germline execution enforces the invariant of \cref{sec:bg-purity} from outside the tool: instead of recognising
hidden code, it arranges each call so that the only execution able to reach a destination never received the
items that host may not see. As in Weismann's germ-plasm theory~\cite{weismann}, nothing an execution acquires
flows back to the process from which later executions start. At install time, the agent platform verifies the signature and starts the release in a
\emph{germline}, a confined process that loads the code and runs its initialisation but never receives user data.
For each call, \Plan{} decides which items each execution holds, and the germline forks one sandboxed \emph{copy}
per \Plan{} group (\cref{fig:arch}c). A trusted broker delivers each request only from the copy that serves its
destination and compares the other copies' requests with it.

We use the notation of \cref{sec:formal-model}: item $x$ may flow to destination or consumer $d$ only if
$\lab(x)\subseteq\clr(d)$. A call (\emph{invocation}) to $f$ carries public metadata $\meta$, items $\Items$,
consumers $\Cons$ (typically the agent) and the destinations $\Decl_f$ declared in the signed manifest; a
\emph{session} is a sequence of calls. The \emph{view} $\View(d)$ is the set of current items $d$ may see; the
\emph{horizon} $\hor$ of a copy, the intersection of its group members' clearances, bounds its state slot
$\slot{\hor}$ and the replies it may receive. A \emph{trigger} is an online request mismatch, and a \emph{verified witness} a trigger that offline replay
attributes to items the destination may not see. Views and horizons can differ: two destinations of a public-only
call, \fn{ocr.example} cleared for \emph{finance} and \fn{mail.example} for \emph{projx}, share a view, but a
reply from \fn{ocr.example} may carry finance data, so \Plan{} gives them separate copies and slots.

\subsection{Plan and declared-destination splitting}\label{sec:design-plan}

\Plan{} reads metadata, labels, clearances, $\Decl_f$ and $\Cons$, never item contents. It groups the destinations and consumers in $\Decl_f\cup\Cons$ by view, then splits any group whose destinations differ in
clearance, so that every reply a copy receives is cleared for all members of its group. A consumer joins a group of
its view whose host clearances all lie inside its own, or else gets a group of its own. Groups run smallest view
first, so the requests delivered from every smaller view are recorded before a copy holding more items runs. In
\cref{fig:arch}c, \fn{ocr.example} is cleared for the invoice's \emph{finance} label and the agent also for the
salary sheet's \emph{hr}, so \Plan{} forms two groups; in most calls of our workloads all members fall into one
group, and the call needs a single copy.

Splitting on \emph{declared} rather than contacted hosts is essential: under lazy splitting, the copy holding
uncleared items decides whether $d$ receives any request, which discloses one bit per call and host
(\cref{lem:existence}, \cref{sec:eval-adaptive}). With declared-destination splitting, which copies exist never
depends on secret data, and within that class \Plan{} uses the fewest copies (\cref{thm:copies}).

\subsection{One call end to end}\label{sec:design-walkthrough}

We follow the \fn{ocr\_annotate} call of \cref{fig:arch}c. \Plan{} returns two groups in order:
$\{\fn{ocr.example}\}$ with view $\{\text{invoice}\}$ and horizon \emph{finance}, then $\{\text{agent}\}$ with view
$\{\text{invoice},\text{salary sheet}\}$ and horizon $\{\emph{finance},\emph{hr}\}$. Copy~1 is forked from the
germline and receives the metadata, the invoice, the salary sheet as a name with empty contents, the twin inputs
(a frozen clock and seed) and the finance slot with its label. The broker delivers its request to \fn{ocr.example}, records it with the reply
as the first request to that host, and returns the reply. Copy~1 serves no consumer, so the broker discards its
return value and kills it with all its descendants.

Copy~2 is forked next and receives both items and the slot of its own horizon. Its requests to \fn{ocr.example} are
never delivered; the broker compares its $k$-th request to that host with the $k$-th request delivered for copy~1.
If they are equal, the broker replays the recorded reply, which is admissible because \fn{ocr.example} is cleared
only for finance; if they differ, or copy~2 sends more or fewer requests, it records a trigger and answers with the
placeholder $\placeholder$. Copy~2 returns its value to the agent, labelled with the join of its view's labels, the
clearances of hosts whose replies reached it and its slot's label, and is killed. Offline, probe replay checks
whether a trigger stems from the salary sheet (\cref{sec:design-witness}).

\subsection{The germline call}\label{sec:design-call}

\begin{algorithm}[t]
  \caption{\Plan{} and the germline call. A host $e$ is \emph{comparable} for group $g$ if its group ran before
  $g$ and $\View(e)\subseteq\View(g)$; it is \emph{admissible} for $g$ if $\clr(e)\subseteq\clr(d)$ for all $d\in g$;
  $\lab(\sigma)$ joins the labels stored in slot $\sigma$ ($\emptyset$ if empty). The tripwire also records mismatches with
  an equal-view host; probe replay always classifies these as state or response divergences.}
  \label{alg:call}
  \footnotesize
  \begin{algorithmic}[1]
    \Procedure{Plan}{$f,\meta,\lab,\Decl_f,\Cons$} \Comment{never reads contents}
      \State $\mathcal{G}\gets$ group $\Decl_f\cup\Cons$ by view $\View(d)$
      \State split each $S\in\mathcal{G}$ until $\clr(e)\subseteq\clr(d)$ for hosts $e$, members $d\in S$
      \State \Return $\mathcal{G}$, ordered so that $\View(g)\subsetneq\View(g')$ puts $g$ first
    \EndProcedure
    \Procedure{GermlineCall}{$\rel,f,\meta,\Items,\Decl_f,\Cons$}
      \State $\twin\gets$ (frozen clock, seed); $\mathit{Rel}[e]\gets\langle\rangle$ for $e\in\Decl_f$
      \For{$g\in\Plan(f,\meta,\lab(\Items),\Decl_f,\Cons)$}
        \If{\gchk{1}, \gchk{2} or \gchk{3} fails for the germline of $\rel$} refuse $\rel$ \EndIf
        \State fork copy in fresh net/IPC/PID namespaces; Landlock, no writes
        \State send $\meta$, $\twin$, whole slot $\sigma=\slot{\hor(g)}$, contents of $\View(g)$
        \Statex \hspace{\algorithmicindent}\hspace{\algorithmicindent}(other items: name only)
        \State $L\gets\lab(\View(g))\join\lab(\sigma)$; \ $n[e]\gets0$ for $e\in\Decl_f$
        \While{copy runs, on message $\mathit{msg}$ within budget $T$}
          \If{$\mathit{msg}$ is a request $q$ to $e\in g$}
            \State deliver $q$, get reply $\rho$; append $(q,\rho)$ to $\mathit{Rel}[e]$
            \State $L\gets L\join\clr(e)$; return $\rho$
          \ElsIf{$\mathit{msg}$ is a request $q$ to $e\in\Decl_f\setminus g$} \Comment{tripwire}
            \State $n[e]\gets n[e]+1$; $(q',\rho')\gets$ $n[e]$-th entry of $\mathit{Rel}[e]$ ($\bot$ if none)
            \If{$e$ admissible for $g$ and $\mathit{Rel}[e]\neq\langle\rangle$} $L\gets L\join\clr(e)$ \EndIf
            \If{$q=q'$ and $e$ admissible for $g$} return $\rho'$ \EndIf
            \If{$q\neq q'$ and $e$ comparable} record trigger $(g,e,q)$ \EndIf
            \State return placeholder $\placeholder$
          \ElsIf{$\mathit{msg}$ is a request to $e\notin\Decl_f$} deny
          \ElsIf{$\mathit{msg}$ is a state write $(\mathit{key},v)$} $\sigma[\mathit{key}]\gets(v,L)$
          \EndIf
        \EndWhile
        \If{deadline reached} kill copy and descendants; output error to consumers in $g$; \textbf{continue} with next group \EndIf
        \For{comparable $e$ with $\View(e)\neq\View(g)$ and $n[e]<|\mathit{Rel}[e]|$}
          \State record trigger $(g,e,\text{absent})$
        \EndFor
        \State kill the copy and all its descendants
        \For{consumer $c\in g$} output the copy's value with label $L$ \EndFor
      \EndFor
    \EndProcedure
  \end{algorithmic}
\end{algorithm}

\Cref{alg:call} gives the procedure. Before every fork, the broker checks that the germline is still data-free:
it has a single thread (\gchk{1}), no descriptor except its control socket, \fn{/dev/null} streams and read-only
allowed files (\gchk{2}), and no shared writable mapping (\gchk{3}); otherwise an inherited pipe or mapping could
carry bytes between copies, so a release that fails a check is refused. Copies run one at a time, smallest view
first, which reduces the parallel-copy timing channel we measured (\cref{sec:scope}). A copy that exceeds its time
budget is killed and its consumers receive an error, so it cannot suppress later groups or delay destinations
served by earlier copies; only groups with equal or incomparable views may wait up to the budget, which is timing
(\asm{1}).

A request thus reaches a destination only from the copy of that destination's group. Hidden code that adds
uncleared items to a request either runs in a copy that has nothing to add or produces a request that is never
delivered; in a call with several copies, that request, or an extra, missing, repeated or reordered one, is
recorded as a trigger (\cref{thm:witness}(c)) at the cost of one digest comparison per request.

\subsection{Per-horizon state slots and labels}\label{sec:design-slots}

Since copies are discarded, flow-compliant tools keep state across calls, such as a session token or a cache,
through the broker. A copy reads and writes only the slot $\slot{\hor}$ of its horizon, much as a browser keeps
cookies per site, so state reaches a later call only if every member of the reading group is cleared for it
(\cref{lem:slot}). Each slot entry stores the label of everything the writing copy had seen (its view, its slot and
the hosts whose replies reached it), and a copy that receives a non-empty slot joins that label into its output
label. For example, if call~1 writes an HR item into the agent's horizon slot, the entry is stored with label
$\{\emph{hr}\}$; a later public-only call with the same horizon receives the whole slot and returns its value
with $L=\{\emph{hr}\}$, which a downstream CaMeL- or FIDES-style policy refuses to forward to a public sink
(\cref{tab:regression}).

Destination hosts are stateful too: a host $e$ may remember anything it was cleared to receive, so its reply
carries the worst-case label $\clr(e)$ and is \emph{admissible} for a copy only if $\clr(e)$ lies inside the
clearance of every member of the copy's group. Replies from remote MCP servers follow the same rule, so a reply
that re-emits earlier data still carries a correct label (\cref{cor:labels}); such tools cannot be forked and
receive prevention but no witnesses (\cref{sec:scope}).

\subsection{Twin execution, witnesses and probe replay}\label{sec:design-twin}\label{sec:design-witness}

A comparison is evidence only if a flow-compliant release produces identical requests in every copy. The broker
therefore fixes twin inputs for each call, chosen without looking at item contents, from which every copy derives
its clocks, exposed performance counters, random generators, \fn{os.urandom} and numeric-library seeds. The shim
works at the Python level, so native code and the CPU cycle counter lie outside it; they affect evidence only.

A trigger is not yet evidence, because the comparing copy may also differ from the copy $g$ that serves $d$ in its
slot or its replies. The \emph{probe} therefore re-runs the release with $g$'s slot, metadata, twin inputs and
replies, so that it differs from $g$ only in the items $d$ may not see, and the \emph{control} re-runs $g$'s exact
inputs (\cref{def:probe}). If the probe's request sequence to $d$ differs on each of $N$ tries while the control
never does, the result is an \emph{item witness}, which, for a release deterministic within the replay model,
proves a flow violation (\cref{thm:witness}). A trigger that disappears in the probe is a state or response
divergence (\cref{prop:state}), and a differing control marks nondeterminism beyond twin control
(\cref{prop:nondet}; \cref{tab:replay}). A verifier holding the release and the named items can repeat probe and
control offline from the witness record. A witness must be handled at the clearance of its contents; its hash
authenticates the code, not the supplied context, and it is a counterexample to flow compliance, not proof of a
past disclosure.

\para{Dampening policy.}\label{sec:design-dampening}
Evidence feeds a local trust policy, separate from prevention, modelled on BGP route-flap damping~\cite{rfc2439}: a release's penalty decays
with a half-life of $\hl=50$ calls to it, refused calls included, grows by $w=250$ per unverified event, and
suppresses the release between thresholds $S=2000$ and $R=750$. A verified witness from a release whose signed manifest
declares determinism adds 2000 and suppresses it at once. No guarantee of \cref{sec:formal} depends on these parameters;
\cref{app:dampening} gives the rule and its analysis.

\section{Formal Model and Guarantees}\label{sec:formal}

\subsection{System model}\label{sec:formal-model}

\begin{definition}[Labels, clearances, flow]\label{def:labels}
Each item $x$ has a label $\lab(x)\subseteq\Tags$ and each destination host or consumer $d$ a clearance $\clr(d)\subseteq\Tags$, both public policy inputs; $x$ may flow to $d$ iff $\lab(x)\subseteq\clr(d)$, and $\join$ is set union.
\end{definition}

\begin{definition}[Release, call, view]\label{def:call}\label{def:view}
A release $\rel$ is a signed package identified by its hash. A call $\call=(\rel,f,\meta_\call,\Items_\call,\Decl_f,\Cons_\call)$ has function $f$, public metadata $\meta_\call$ (function name, arguments, item names and labels), labelled items $\Items_\call$, the destination hosts $\Decl_f$ declared for $f$ in the signed manifest, and consumers $\Cons_\call$ of the return value.
The view of $d$ is $\View_\call(d)=\{x\in\Items_\call:\lab(x)\subseteq\clr(d)\}$; we drop the subscript $\call$ when the call is clear.
\end{definition}

\begin{definition}[Horizon, state slots, twin inputs]\label{def:horizon}\label{def:twin}
A copy serves one \Plan{} group $g\subseteq\Decl_f\cup\Cons_\call$ whose members share one view $\View(g)$, and has horizon $\hor(g)=\bigcap_{d\in g}\clr(d)$; $\hor_\call(d)$ is the horizon of the copy serving $d$.
The broker delivers the whole slot $\slot{\hor(g)}$ when the copy starts and accepts writes only to it; each entry is stored with the join of the writer's view labels, its slot's label and the clearances of the hosts whose replies reached it, and $\lab(\sigma)$ joins a slot's entry labels ($\emptyset$ if empty).
All copies of the call receive twin inputs $\twin_\call=(t_\call,s_\call)$, a frozen clock and a seed chosen without reading item contents.
\end{definition}

\begin{definition}[Responses and transcripts]\label{def:responses}\label{def:transcript}
A reply from destination host $e$ carries the worst-case label $\clr(e)$ and is \emph{admissible} for a copy serving $g$ if $\clr(e)\subseteq\clr(d)$ for all $d\in g$.
Under the \emph{positional rule}, a copy outside $e$'s group whose $j$-th request to $e$ matches the $j$-th delivered one by digest receives the recorded reply if it is admissible, and the placeholder $\placeholder$ otherwise.
The transcript $T_\call(d)$ lists the requests and replies of the copy serving $d$, and $\Resp_\call(d)$ is the resulting reply environment.
\end{definition}

\begin{lemma}[Slot invariant]\label{lem:slot}
Every value in $\slot{\hor}$ is computed only from items labelled inside $\hor$, earlier contents of $\slot{\hor}$, public metadata, twin inputs and replies from destination hosts $e$ with $\clr(e)\subseteq\hor$, and its stored label lies inside $\hor$.
\end{lemma}

\begin{definition}[Sink noninterference]\label{def:sni}
Following Goguen and Meseguer~\cite{goguen} with each destination host as a sink, release $\rel$ satisfies sink noninterference toward $d$ if, in every call $\call$, its request sequence to $d$ is a function of five arguments only:
\begin{equation}\label{eq:sni}
\Req_\call(\rel,d)=\Fr\big(\meta_\call,\ \restr{\Items_\call}{\View_\call(d)},\ \slot{\hor_\call(d)},\ \twin_\call,\ \Resp_\call(d)\big),
\end{equation}
where $\restr{\Items_\call}{\View_\call(d)}$ denotes the contents of the items $d$ may see, and the $j$-th request may depend on the replies $\Resp_\call(d)$ gives to earlier ones.
A \emph{flow violation} is a pair of executions that agree on all five arguments but send different request sequences to $d$; a release with none toward any destination host is \emph{flow-compliant}.
\end{definition}

Flow compliance is a property of a release; the mechanism enforces the stronger, release-independent property that whatever $d$ \emph{receives} satisfies \cref{eq:sni}.

\subsection{Assumptions}\label{sec:formal-assumptions}

Against the adversary of \cref{sec:threat-adversary}, the theorems rely on six assumptions, enforced as listed in \cref{tab:assumptions}.

\begin{description}
\item[\asm{1} Confinement and termination.] A copy's only external channel is its broker socket; copies run one at a time, smallest view first, each ending with its descendants before the next starts. A copy that exceeds a total time budget $T$ is killed, its consumers receive an error, and the other copies run as planned. Whether and how often a request is sent lies inside the model; when it is sent, including where the budget cuts a sequence, is timing and lies outside.
\item[\asm{2} Data-free germline.] At every fork the germline passes \gchk{1}--\gchk{3} (\cref{sec:design-call}), and its image $\Img$ was built before any user data existed and depends only on $\rel$.
\item[\asm{3} Trusted base.] The kernel, broker, label policy and signature verification are correct, and the broker delivers each request to its named destination only (no redirects).
\item[\asm{4} Twin scope.] A Python-level shim controls the clock, seed, \fn{os.urandom} and performance counters, not the CPU cycle counter. Only \cref{thm:witness} uses \asm{4}.
\item[\asm{5} Public metadata.] Metadata is cleared for every destination; secret content travels as a labelled item.
\item[\asm{6} Responses.] A reply from $e$ is data labelled $\clr(e)$, served under the rule of \cref{def:responses}.
\end{description}

The prototype enforces \asm{1}, \asm{2}, \asm{6} and the probe through obligations \obl{1}--\obl{5} and the per-copy deadline; property and regression tests (\cref{tab:obligations,tab:regression}) support the correspondence between model and prototype for the behaviours they exercise, not for every machine-level behaviour. \Cref{sec:impl} states two remaining prototype gaps (\cref{tab:boundary}).

\subsection{Sink noninterference}\label{sec:formal-sni}

\begin{theorem}[Sink noninterference]\label{thm:sni}
Assume \asm{1}--\asm{3}, \asm{5} and \asm{6}.
For every release $\rel$, call $\call$ and declared destination host $d\in\Decl_f$, the request sequence $\Deliv_\call(d)$ delivered to $d$ satisfies \cref{eq:sni} in place of $\Req_\call(\rel,d)$, and hence does not depend on the contents of any item outside $\View_\call(d)$.
If $\rel$ uses randomness the broker does not control, the distribution of delivered sequences is such a function.
\end{theorem}

\begin{corollary}[Sessions]\label{cor:sessions}
Across a session, what destination host $d$ receives depends only on items labelled inside $\clr(d)$, from this and earlier calls, and on admissible replies.
\end{corollary}

\begin{corollary}[Sound output labels]\label{cor:labels}
For a group $g$ with consumer $u$, let $\sigma_g$ be the slot delivered to $g$'s copy and $E_g$ the destination hosts whose reply reached it, live or as an admissible recorded reply, together with every admissible host with a non-empty record to which it sent a request. The returned value carries
\begin{equation}\label{eq:label}
\lab(g)=\bigsqcup_{x\in\View(g)}\lab(x)\ \join\ \bigsqcup_{e\in E_g}\clr(e)\ \join\ \lab(\sigma_g),
\end{equation}
which lies inside $\clr(u)$ and contains the label of every input of the copy: its items, slot entries, replies, and whether a request matched a recorded one.
\end{corollary}

\Cref{cor:sessions} covers carry-over attacks such as \atk{pool}, and \cref{cor:labels} gives downstream policy engines labels that cover stored state.

\subsection{Witnesses and detection coverage}\label{sec:formal-witness}

\begin{definition}[Probe replay, control, item witness]\label{def:probe}\label{def:control}\label{def:witness}
For a trigger between a comparing copy with view $W\supsetneq\View(d)$ and $d$'s group $g$, the \emph{probe} runs $\rel$ on $\meta$, the contents of $W$, $g$'s slot $\slot{\hor(d)}$ and $\twin$ under $\Resp(d)$, so it differs from $g$ only in the contents of $W\setminus\View(d)$; the \emph{control} does the same with $g$'s item contents.
An \emph{item witness} for $(\rel,d)$ is a probe whose request sequence to $d$ differs from the delivered one on each of $N\ge1$ tries while the control never differs; it records the release hash, $d$, the item names of both views, the slot snapshot, $T(d)$ and both sequences.
\end{definition}

\begin{theorem}[Witnesses and detection coverage]\label{thm:witness}
Assume \asm{1}--\asm{6}; a release is \emph{deterministic within the replay model} if it draws on no source outside \asm{4}.
(a)~\emph{Witness soundness:} if $\rel$ is deterministic within the replay model and an item witness for $(\rel,d)$ exists, $\rel$ has a flow violation toward destination host $d$.
(b)~\emph{No false item witnesses:} if such a $\rel$ is flow-compliant toward $d$, stateful or not, no item witness for $d$ is produced.
(c)~\emph{Detection coverage:} in a call in which \Plan{} forks more than one copy, if a copy whose view strictly contains $\View(d)$ finishes within its deadline and its request sequence to $d$ differs from the delivered sequence in content, count or order, including a request absent from either, the tripwire records a trigger, and probe replay classifies it as an item witness, a state or response divergence (\cref{prop:state}) or nondeterminism outside \asm{4} (\cref{prop:nondet}).
\end{theorem}

Parts (a) and (b) hold because the probe and $g$ evaluate one function on arguments that differ only in uncleared contents.
Part~(c) does not cover single-copy calls, where nothing is compared, return values, timing, undeclared hosts, which are denied rather than compared, or copies whose view does not strictly contain $\View(d)$; a divergence caused only by sources outside \asm{4} triggers but is never verified.

\subsection{Copies per call}\label{sec:formal-copies}

\begin{definition}[Data-independent fork rule, transparency]\label{def:forkrule}\label{def:transparency}
A fork rule is \emph{data-independent} if it fixes, before any run, the runs and their item sets and the rule that selects which run's output each member of $\Decl_f\cup\Cons$ receives, as a function of $f$, $\meta$, labels, clearances, $\Decl_f$ and $\Cons$ only.
A mechanism is \emph{transparent} if, for every flow-compliant release, each destination host and consumer receives exactly what that release produces on its own view with admissible replies.
\end{definition}

\begin{theorem}[Lower bound on copies]\label{thm:copies}
Call a mechanism \emph{black-box} if it learns a release only by running it and delivers only outputs of runs on item subsets, in which other items arrive as empty names.
Consider black-box mechanisms with a data-independent fork rule; releases with one manifest then receive the same runs and the same output rule.
(a)~Any such mechanism that satisfies \cref{thm:sni} for every release and is transparent must, in every call, run the release on each distinct view among $\Decl_f\cup\Cons$, and within a view once per distinct clearance among its destination hosts, plus once, if there are any, for the consumers whose clearance contains no host clearance of that view.
(b)~\Plan{} creates exactly these runs, so declared-destination splitting is optimal within this class.
\end{theorem}

The class matters: under \emph{lazy splitting}, a fork rule that is not data-independent, $d$'s copy is forked only when the full-view copy contacts $d$.

\begin{lemma}[A data-dependent rule discloses one bit]\label{lem:existence}
Under lazy splitting, whether destination host $d$ receives any request in a call is a function of items outside $\View(d)$, one bit per call and destination host, which the tripwire cannot see because the delivered request comes from the copy with view $\View(d)$.
\end{lemma}

The price of a data-independent rule is that the bound counts declared rather than contacted destinations; the existence-bit test measures the bit (\cref{sec:eval-adaptive}), and \cref{app:proofs} gives the cost of a call.

\section{Implementation}\label{sec:impl}

The \sys{} prototype realises \cref{alg:call} on stock Linux without kernel changes, in Python~3.13. Its core has
999 lines (broker with \Plan{} and baselines, germline fork server, confinement, witnesses, dampening); the tool
host inside every copy adds 151 lines and the germline for unmodified MCP servers 75. For the comparison with
MCPGuard (\cref{sec:eval-mcpguard}), the broker also runs inside MCPGuard's own runner. A publisher Ed25519
key~\cite{ed25519} signs TUF-style metadata~\cite{tuf} with each release's version, SHA-256 hash and length; the
host pins the key and verifies every release, including updates, before extracting it. The signed manifest lists
readable paths, destinations and whether the release declares determinism. Other language
runtimes would use snapshot-based germlines.

\para{Germline and copies.} The germline imports the package and runs its initialisation under a read-only Landlock
ruleset~\cite{landlock} with standard streams bound to \fn{/dev/null}, and the broker inspects it through
\fn{/proc} for \gchk{1}--\gchk{3} before every fork. Each copy is PID~1 of its own PID namespace, with fresh
network and IPC namespaces and a further read-only Landlock layer. Its only channel is a \fn{socketpair} carrying length-prefixed JSON, over which it receives the public
metadata, its view's items as \fn{memfd} files, its whole horizon slot and the twin inputs. The tool host then
freezes wall-clock time and \fn{datetime}, replaces the monotonic clock and performance counters with a
deterministic counter, and seeds \fn{random}, \fn{os.urandom} and \fn{numpy.random}. An optional pool keeps
pre-forked, data-free copies ready. For an MCP server, each copy instead continues the server's own stdio loop on a
fresh socket, over which the host speaks MCP JSON-RPC, and relays HTTP to the broker over a second socket; servers
run without code changes, although git needs a read-write \fn{/dev/null} (\cref{sec:eval-real}).

\para{Broker.} A copy has no network: the tool host relays its HTTP requests over the channel, and the broker
issues every request itself, only for the copy that holds the destination's view, so exactly one copy delivers
each external effect. Comparison, positional replay, slot labels and probe replay follow \cref{alg:call}; the
trigger fires at the first mismatch or when a copy ends with a shorter sequence. Each copy has an absolute budget
of 30\,s from its fork, kept on the broker's clock, so neither the copy's messages nor its frozen clock extend it,
and the broker's own network waits end with the budget. A copy that exceeds it, crashes or sends a malformed message
is killed with its descendants, its result becomes an error, and the next call proceeds normally. The broker does
not follow redirects: a 3xx reply returns to the copy as the destination host's reply.

\para{Enforcement boundary.} Each obligation maps to one mechanism and one property test, and regression tests
cover slot labels, sequence comparison, the deadline and redirects (\cref{tab:obligations,tab:regression}).
\Cref{tab:boundary} states where enforcement ends; the entries in its right column lie outside the model.

\begin{table}[t]
  \centering\footnotesize
  \setlength{\tabcolsep}{3pt}
  \caption{Enforcement boundary of the prototype.}
  \label{tab:boundary}
  \begin{tabular}{@{}>{\raggedright\arraybackslash}p{0.17\columnwidth}
                  >{\raggedright\arraybackslash}p{0.42\columnwidth}
                  >{\raggedright\arraybackslash}p{0.36\columnwidth}@{}}
    \toprule
    Aspect & Enforced by \sys{} & Outside / required \\
    \midrule
    Init reads & Read-only allowlist, no network & Allowlist holds no user data \\
    Descriptors & \gchk{2}: control socket, \fn{/dev/null}, read-only files & Anything else: release refused \\
    Messages & Three types; broker picks slot, computes labels & No size quotas \\
    Clocks, RNG & Python-level shim (\asm{4}) & Native code weakens evidence only \\
    Slots & Per release and horizon; an update has its own & One user per deployment \\
    Destinations & URL host name, compared literally; redirects not followed & DNS and TLS attacks out of scope \\
    Termination & Absolute 30\,s budget per copy on the broker's clock & No memory quota \\
    \bottomrule
  \end{tabular}
\end{table}

\section{Evaluation}\label{sec:eval}

We evaluate germline execution against signed releases that perform their advertised function while carrying hidden code, through five research questions: whether a signed tool still discloses data under MCPGuard-Dynamic and whether germline execution prevents it in the same harness (RQ1, \cref{sec:eval-mcpguard}); prevention across vectors, compositions, generated variants and stateful remote tools (RQ2, \cref{sec:eval-matrix}); utility for flow-compliant tools, unmodified servers, stateful workflows and an LLM-driven agent (RQ3, \cref{sec:eval-utility}); soundness and coverage of witnesses, including against adaptive releases (RQ4, \cref{sec:eval-witness}); and latency and memory cost (RQ5, \cref{sec:eval-overhead}).
\Cref{app:eid} maps each experiment to its question and result file.

\subsection{Setup and methodology}\label{sec:eval-setup}

\para{Testbed and counts.}
Five MCP servers (documents, SQL, web, lint, mail) each ship a flow-compliant release 1.0.0 and signed updates with hidden code, 52 in total, implementing the 15 attack variants of \cref{sec:threat-vectors}; with two benchmark releases, all 59 signed releases verify against the pinned key and complete their advertised function.
We wrote these 59 releases by hand; the generated variants below are a separate corpus. By design, the test world is one user's home directory, small enough to check every flow: PDFs labelled \textit{hr}, \textit{board}, \textit{finance}, \textit{projx} and public, two SQLite databases, a code repository, mail drafts and credential files, each protected item carrying a fresh random canary per trial.
The scripted matrix has 51 attack cases (the two cooperating \atk{xtool} releases form one) in 3 trials, or 153 \emph{attack runs} per runtime, plus 33 benign calls per trial; the single-harness MCPGuard comparison has 5 variants, 5 runtimes and 3 trials (75 runs), and the LLM-driven agent 7 tasks, 6 releases, 4 runtimes and 3 repetitions (504 episodes).

\para{Disclosure oracle.}
Five HTTP sinks record every byte they receive, and an oracle that runs outside the runtime under test counts a disclosure when a canary, decoded from raw text, URL encoding, nested base64, hex or zero-width characters, reaches a sink whose clearance excludes its label (per recipient for mail).
The tables count every detected disclosure in an attack run; a disclosure that also occurs with the flow-compliant release in the same runtime is not attributed to the tool, which concerns six \rtSig{} runs (\cref{tab:matrix}).
Prevention under \rtGerm{} is structural (\cref{thm:sni} holds for any encoding), whereas the measurements count detected canaries; every zero below is scoped to our corpus and this oracle.

\para{Runtimes.}
\rtSig{} runs one persistent process per server after signature verification and key pinning; \rtIFC{} adds per-call output labels as in FIDES and CaMeL~\cite{fides,camel}; \rtConf{} adds AgentBound/MCPGuard-class confinement~\cite{agentbound,mcpguard} (network namespace, Landlock~\cite{landlock} allowlists, broker egress allowlist); \rtFresh{} also starts a fresh process per call; and the ablation \rtNoSplit{} runs one copy per call holding all inputs.
We did not integrate a snapshot/restore runtime such as Groundhog~\cite{groundhog}: its security effect, a clean state per call without per-destination views, is what \rtNoSplit{} isolates, and germline execution already gains its speed by forking a pre-initialised process, so speed claims are relative to a fresh process per call only.

\para{MCPGuard-Dynamic.}
We built MCPGuard-Dynamic~\cite{mcpguard} from its public repository (commit \fn{825fdc0a}), enabled the BPF LSM and loaded its four eBPF programs, proxy, policy format and AgentBound reproduction.
Our least-privilege policies in its format cover a six-call session that summarises HR and board documents, sends an invoice to the OCR destination (cleared for \textit{finance}) and posts a press-release summary to the public blog.
MCPGuard's runner executes its seven configurations (\cref{tab:mcpguard}(b)) with a positive control, a raw-socket transfer to an unlisted host.
The single harness runs the two strongest, \mcg{C-full} and \mcg{C-full-percall} (per-call policy enforcement), through MCPGuard's proxy and our runtimes through our broker, in one process with the same session, world, canaries and oracle.

\para{Generated variants.}
The generated variants target the \fn{read\_note} tool of MCPGuard's signed note server (commit \fn{825fdc0a}).
We generated 92 distinct packages in five families from 189 attempts with a local Qwen3.5-27B model, signed each and
admitted it only if its output matches the clean tool and it discloses a canary with no defence, which 91 do
(71 send to an attacker-owned host that no policy lists, 20 to a policy-allowed host; protocol in \cref{app:e13}).

Experiments run on a GPU server VM or a 2-vCPU cloud container (\cref{app:eid}); numeric libraries run single-threaded in the germline to satisfy \gchk{1}, and the agent's model runs with reasoning off.

\subsection{RQ1: Matched comparison with MCPGuard-Dynamic}\label{sec:eval-mcpguard}

In MCPGuard's own runner, the eBPF configurations stop the raw-socket variant \atk{direct}, while \atk{pool} and \atk{smuggle} disclose through the authorised OCR destination under all seven configurations (\cref{tab:mcpguard}(b); per-configuration rows in \cref{tab:e8a-full}).
In the single harness, \atk{pool} is the matched result: it stashes items from the HR and board calls and adds them to the next authorised OCR request.
Our own \atk{smuggle} release sent nothing in this session, so its prevention is measured in the scripted matrix (9 of 15 runs disclose under \rtSig{}, 3 under \rtConf{}, none under \rtGerm{}; \cref{tab:e1-variant}); \atk{launder\_text} reaches no sink in MCPGuard's runner even under \mcg{C0}.
\rtGerm{} produces no witness here, because no copy holds carried-over data or items outside its view.

\takeaway{MCPGuard confines destinations as designed, yet the signed \atk{pool} release discloses through the authorised OCR destination in 3 of 3 trials under every configuration; none was detected under germline execution.}

This marks the boundary of destination-based confinement: MCPGuard binds authority to a process and its destinations, and its per-call mode re-applies the policy while the process and its memory persist, whereas germline execution binds authority to the items each copy may see, so the two compose (\cref{sec:discussion-composition}).

\subsection{RQ2: Prevention across vectors, generated variants and remote tools}\label{sec:eval-matrix}

\begin{figure}[t]
  \centering
  \safegraphic[width=\columnwidth]{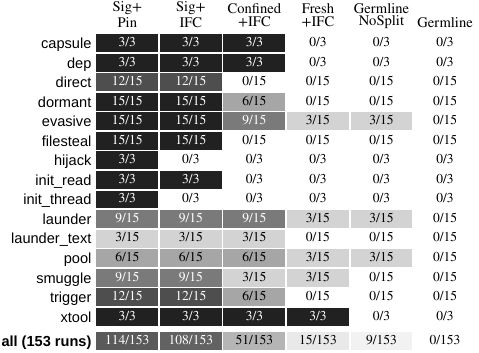}
  \caption{Scripted security matrix: attack runs with a detected canary disclosure per variant and runtime (51 cases, 3 trials; darker cells disclose more often). Totals: 74.5\%, 70.6\%, 33.3\%, 9.8\%, 5.9\% and \textbf{0.0\%}. In 6 \rtSig{} \V{1} runs only the agent forwards protected replies, so 108/153 (70.6\%) are attributable to the tool (\cref{tab:e1-variant}). Every runtime completes 99/99 benign steps.}
  \label{fig:heatmap}\label{tab:matrix}
\end{figure}

No canary disclosure was detected in 153 attack runs under \rtGerm{} (\cref{fig:heatmap}).
Each baseline binds authority to a different object and so misses a different vector: output labels miss carry-over and laundering, and confinement misses disclosure through permitted destinations, content-triggered code (\V{1}+\V{2}) and a second server that ships what the first dropped (\V{1}+\V{3}).
A fresh process per call removes carry-over but still mixes items within a call, and \rtNoSplit{} shows that without split views one copy can place protected items in a request to a less-cleared destination.
The dependency bump \atk{dep} discloses under \rtConf{} but not under \rtGerm{}, and the germline checks refuse the two releases that attack initialisation (\atk{init\_thread}, \atk{hijack}) before their first call.
\rtGerm{} records 12 triggers in attack runs and none in benign runs (\cref{sec:eval-witness}); per-variant disclosure counts are in \cref{tab:e1-variant}.

\para{Generated variants.}\label{sec:eval-scale}
No canary disclosure was detected for any of the 91 validated variants (20 direct, 20 allowed-request, 20 dormant, 17 triggered, 14 carry-over) in three trials.
An egress-allowlist defence of the MCPGuard/AgentBound class prevents the 71 that send to an attacker-owned host and none of the 20 that send to a policy-allowed host, 71 of 91 (78\%) overall; germline execution prevents those 20 at the view boundary, because the copy serving the public destination never receives the protected note, so no witnesses arise.
As an author-generated corpus, it shows that prevention generalises beyond our hand-written payloads; it is not an independent benchmark and does not show how common such code is.

\para{Stateful remote tools and agent labels.}\label{sec:eval-labels}
A stateful remote tool, which cannot be forked, discloses 2 canaries to the public sink under per-call labels and none under the worst-case reply labels \sys{} assigns to remote tools (\cref{sec:design-slots}), which block 3 public posts of a flow-compliant remote tool where per-call labels block 1.
In CaMeL's reference interpreter (commit \fn{f083b6b}) with its AgentDojo Workspace policy~\cite{camel,agentdojo}, run on hand-written plans, a second reply of a stateful tool that re-emits a private email body from the first call is labelled public and passes the \fn{send\_email} policy to an untrusted recipient in NORMAL and STRICT modes, while sending the body directly is denied.
This needs the deployer to list the stateful tool as side-effect-free, a declaration CaMeL takes on trust for exactly the trusted signed component of our threat model. No FIDES code was available to us.
Under \sys{}, a slot-stored HR item recalled by a later public-only call returns labelled \textit{hr} and is refused by a downstream policy, while a public-only control is forwarded (\cref{tab:regression}).

\subsection{RQ3: Utility and compatibility}\label{sec:eval-utility}

A tool call is compatible only if its output equals that of the unprotected server, and an LLM task succeeds
only if the answer contains the expected key string or the requested post was made, as checked automatically.

\begin{table}[tb]
  \centering
  \footnotesize
  \caption{LLM-driven agent (Qwen3-32B-AWQ chooses every tool call): canaries disclosed per repetition (identical in all 3 repetitions) and task success.}
  \label{tab:llm}
  \setlength{\tabcolsep}{3.5pt}
  \begin{tabular}{@{}lccccc@{}}
    \toprule
    Release & \rtSig & \rtIFC & \rtConf & \rtGerm & Tasks \\
    \midrule
    \atk{benign}        & 0  & 0  & 0 & \textbf{0} & 7/7 \\
    \atk{pool}          & 2  & 2  & 2 & \textbf{0} & 7/7 \\
    \atk{dormant}       & 28 & 28 & 0 & \textbf{0} & 7/7 \\
    \atk{launder\_text} & 1  & 1  & 1 & \textbf{0} & 7/7 \\
    \atk{direct}        & 8  & 8  & 0 & \textbf{0} & 7/7 \\
    \atk{smuggle}       & 6  & 6  & 3 & \textbf{0} & 7/7 \\
    \bottomrule
  \end{tabular}
\end{table}

\para{LLM-driven agent.}\label{sec:eval-llm}
Qwen3-32B-AWQ~\cite{qwen3} chooses every tool call in seven document tasks, for the flow-compliant release and
five releases with hidden code; the agent completes every task and no canary disclosure was detected under \sys{}
(\cref{tab:llm}). With
reasoning enabled (Qwen3.5-27B), \atk{launder\_text} disclosed under no runtime, while \atk{pool} still disclosed
2 canaries under \rtIFC{} and \rtConf{} and none under \sys{}.

\para{Unmodified servers and workflows.}\label{sec:eval-real}
We called four unmodified official MCP servers~\cite{mcpspec} 780 times under \sys{}: the two tools of
\fn{mcp-server-time} and the \fn{fetch} tool of \fn{mcp-server-fetch} 200 times each, and three tools each of
\fn{mcp-server-sqlite} and \fn{mcp-server-git} 30 times each. Every call returned the same output as the
unprotected server, apart from the clock value of \fn{get\_current\_time}, while median per-call latency rose from
1.2--18.4\,ms to 39.0--110.3\,ms (\cref{app:tables}). The fetch tool's destination was declared per call from
its URL (\asm{5}). Two workflows also
behave identically, with no witness: a notes tool that keeps its notes through the broker's slot interface, rather than in process memory, adds two notes and shows them; and a
call that sends an invoice (\textit{finance}) to the OCR host and returns a payslip (\textit{hr}) only to the
agent.

\subsection{RQ4: Evidence quality and detection coverage}\label{sec:eval-witness}

We test the three parts of \cref{thm:witness}, each against its reference where one exists
(\cref{tab:obligations,tab:evidence,tab:regression}).

\begin{table}[tb]
  \centering
  \caption{Evidence tests. \emph{Reference} is \sys{} with the named component switched off, or the named ablation. \obl{1}--\obl{5} gave identical results on Landlock ABI~7 and ABI~4.}
  \label{tab:evidence}
  \footnotesize
  \setlength{\tabcolsep}{3pt}
  \begin{tabular}{@{}>{\raggedright\arraybackslash}p{2.35cm}>{\raggedright\arraybackslash}p{2.35cm}%
      >{\raggedright\arraybackslash}p{2.4cm}@{}}
    \toprule
    Test & Reference & \sys{} \\
    \midrule
    4 triggers from 4 releases
      & probe replay off: 4/4 confirmed
      & 3 item witnesses (\atk{pool}, \atk{launder}, \atk{evasive}); 1 state divergence (\atk{capsule}) \\
    Twin test: clock and nonce, 20 runs
      & no twin: 20/20 false witnesses
      & 0/20 \\
    \obl{5}: flow-compliant call counter
      & probe replay off: confirmed (false positive)
      & state divergence, weight 250 \\
    \obl{1}: pipe opened at init
      & release starts
      & refused (\gchk{2}) \\
    \obl{2}: shared mapping at init
      & release starts
      & refused (\gchk{3}) \\
    \obl{3}: child outlives its copy
      & 1 live process
      & 0 \\
    \obl{4}: reply from a \fn{projx} host, \fn{finance} consumer
      & 1 copy; consumer receives the reply
      & 2 copies; consumer receives the placeholder \\
    Existence bit, 64 calls
      & lazy splitting: 64/64 bits, 0 triggers
      & 0 bits (mutual information 0.000); triggers on 23/23 and 41/41 secret-dependent attempts \\
    Deniability, 6 noise variants
      & ---
      & 0 HR items at the OCR host \\
    \bottomrule
  \end{tabular}
\end{table}

\begin{table}[tb]
  \centering
  \caption{Regression tests for slot labels, sequence comparison, the per-copy deadline and redirects, and the two workflows. \emph{Reference} is \sys{} with the named component switched off, or the unprotected server for the workflows.}
  \label{tab:regression}
  \footnotesize
  \setlength{\tabcolsep}{3pt}
  \begin{tabular}{@{}>{\raggedright\arraybackslash}p{2.45cm}>{\raggedright\arraybackslash}p{2.2cm}%
      >{\raggedright\arraybackslash}p{2.45cm}@{}}
    \toprule
    Test & Reference & \sys{} \\
    \midrule
    \multicolumn{3}{@{}l}{\emph{Slot labels} (reference: slot-label join off)} \\
    HR item stored, recalled by a public-only call
      & label $\emptyset$; forwarding to public sink allowed
      & label \{\textit{hr}\}; forwarding refused \\
    Public-only control
      & label $\emptyset$; allowed
      & label $\emptyset$; allowed \\
    Mismatched request to an admissible host with a record
      & label \{\textit{hr}\}; host clearance missing
      & label \{\textit{finance}, \textit{hr}\} \\
    \midrule
    \multicolumn{3}{@{}l}{\emph{Sequence comparison} (reference: comparison off)} \\
    Extra request (empty delivered sequence)
      & 0 triggers
      & 1 trigger; verified item witness \\
    Suppressed request
      & 0 triggers
      & 1 trigger; verified item witness \\
    Duplicated request
      & 0 triggers
      & 1 trigger; verified item witness \\
    Reordered requests
      & 0 triggers
      & 1 trigger; verified item witness \\
    Flow-compliant repeated identical request
      & 0 triggers
      & 0 triggers \\
    \midrule
    \multicolumn{3}{@{}l}{\emph{Per-copy deadline} (test setting 3\,s; default 30\,s)} \\
    Copy that never finishes
      & ---
      & killed at 3.0\,s; call returns an error; next call succeeds \\
    Copy that keeps sending messages
      & ---
      & killed at 3.0\,s; next call succeeds \\
    Copy waiting on a slow destination host
      & ---
      & killed at 3.0\,s; next call succeeds \\
    \midrule
    \multicolumn{3}{@{}l}{\emph{Redirects} (reference: broker follows redirects)} \\
    302 to attacker host
      & 1 request at attacker host
      & 0 requests; copy receives the 302 \\
    307 to attacker host (body kept)
      & 1 request, 21 item bytes at attacker host
      & 0 requests; copy receives the 307 \\
    \midrule
    \multicolumn{3}{@{}l}{\emph{Workflows} (reference: unprotected server)} \\
    Notes: add, add, show
      & ``first note. second note.'', label $\emptyset$
      & identical \\
    Invoice to OCR host, payslip to agent
      & agent value and OCR request
      & identical; 0 witnesses \\
    \bottomrule
  \end{tabular}
\end{table}

\para{Verified witnesses.}
\sys{} raised 12 triggers in the scripted matrix, one per trial for four document-server releases. The triggers
of \atk{pool}, \atk{launder} and \atk{evasive} reproduce in every probe while the control stays constant, so each
yields an item witness and, by \cref{thm:witness}(a), a verified flow violation. The \atk{capsule} trigger vanishes
with the destination's slot, a state divergence (\cref{prop:state}); its disclosure was prevented.
A witness occupies 7,450 bytes and verifies in about 0.35\,s, off the agent's critical path.

\para{No false item witnesses.}
A flow-compliant tool that reads the clock and draws a nonce yields a false witness in all 20 runs without twin
execution and in none with it. A flow-compliant per-horizon call counter fires the tripwire once, and the probe,
replaying with the destination's slot, classes it as a state divergence, as \cref{thm:witness}(b) requires.

\para{Detection coverage.}
Four fixtures exercise \cref{thm:witness}(c): the copy holding an uncleared item adds a request, omits one,
duplicates one, or reorders two. Each raises one trigger that probe replay verifies as an item witness, and none
fires with sequence comparison switched off; a flow-compliant tool that repeats an identical request raises no
trigger.

\para{Existence bit.}\label{sec:eval-adaptive}
To test \cref{lem:existence}, a release sends the cleared invoice to the OCR host in call $i$ only if bit $i$ of a
64-bit secret is one (23 ones); a copy that cannot see the secret falls back to sending or to not sending. With the
sending fallback, lazy splitting lets the host recover all 64 bits (0.942 bits per call) without a trigger; with the
other, it recovers 41 bits, no more than always guessing zero. Declared-destination splitting discloses 0.000 bits
in both settings, and its tripwire fires on every secret-dependent attempt (23/23 added and 41/41 omitted
requests).

\para{Deniability, pre-deployment checks and dampening.}\label{sec:eval-dampening}
Six variants that add noise from \fn{os.urandom}, the performance counter or the clock disclosed no HR content,
and their secret-dependent triggers verified in 20 of 20 calls. Of the 52 signed releases with hidden code, a static scan flags only 1 after obfuscation, and a three-call honey test
flags 28, missing releases of \atk{dormant}, \atk{trigger}, \atk{pool} and \atk{smuggle} among others. Dampening
suppresses an always-diverging release after 9 calls; in simulation, random unexplained divergence on 3\% of calls
causes false suppression within 10,000 calls with probability 2.4\%, and at 1\% none within 100,000 calls, while
bursty divergence is suppressed more often (\cref{fig:eval}(d), \cref{app:dampening}).

\subsection{RQ5: Cost}\label{sec:eval-overhead}

\begin{figure*}[t]
  \centering
  \safegraphic[width=\textwidth]{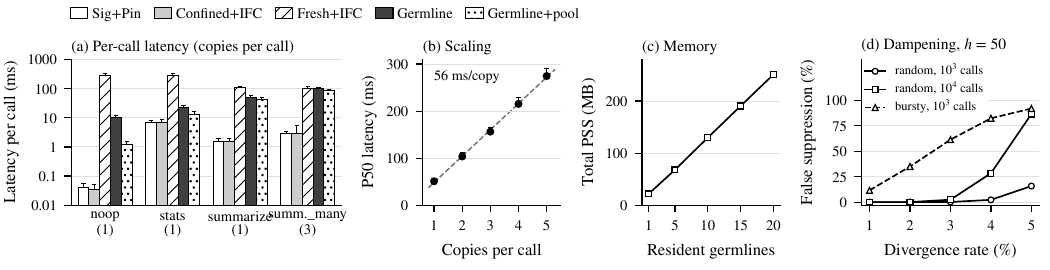}
  \caption{Cost and dampening. (a)~Per-call latency on the server (log scale; P50 with whisker to P99; copies per call in parentheses). Fresh+IFC starts a fresh process per call; Germline+pool forks copies ahead of time. (b)~P50 latency grows linearly with copies per call (least-squares slope 56\,ms per copy). (c)~Total memory (PSS) of resident germlines. (d)~Probability that dampening suppresses a release whose unexplained divergence occurs at the given rate, at the operating point $w=250$, $S=2000$, $h=50$ ($p^{*}=11.0\%$; Monte-Carlo, \cref{tab:montecarlo}); bursty schedules concentrate the same rate into bursts of 10 calls.}
  \label{fig:eval}
\end{figure*}

\begin{table}[tb]
  \centering
  \footnotesize
  \caption{Median per-call latency of \sys{} (P50, median of three passes), time added to \rtSig{} and
    \rtConf{}, and time saved and speed-up against a fresh process per call (ms). The 2--28$\times$ speed-up holds
    for the single-copy cases only.}
  \label{tab:overhead}
  \setlength{\tabcolsep}{2.5pt}
  \begin{tabular}{@{}lcrrrrr@{}}
    \toprule
    Case & \shortstack{Co-\\pies} & P50 & \shortstack{+ms vs\\\rtSig{}} & \shortstack{+ms vs\\Conf.} &
      \shortstack{saved vs\\fresh} & \shortstack{speed-\\up} \\
    \midrule
    \fn{noop}            & 1 & 9.9   & 9.9  & 9.9  & 272.2 & 28.4$\times$ \\
    \fn{stats}           & 1 & 21.8  & 14.9 & 14.9 & 268.0 & 13.3$\times$ \\
    \fn{summarize}       & 1 & 48.5  & 47.0 & 47.0 & 58.4  & 2.2$\times$ \\
    \fn{summarize\_many} & 3 & 97.3  & 94.5 & 94.5 & 5.5   & 1.1$\times$ \\
    \bottomrule
  \end{tabular}
\end{table}

\para{Per-call latency.}
A call served by one copy takes 10--49\,ms, 2--28$\times$ faster than a fresh process per call
(\cref{tab:overhead}, \cref{fig:eval}(a)); for summarisation, most of the added time is the PDF parse repeated inside
the copy. P99 stays within 7\,ms of P50, and a pool of pre-forked copies saves a further 7--9\,ms.

\para{Multiple copies and scaling.}
Copies run in sequence, so a call pays for every copy it needs: \fn{summarize\_many} runs three copies in
97.3\,ms, barely faster than a fresh process (102.8\,ms), and latency grows linearly with copies
(\cref{fig:eval}(b)); external effects are never duplicated. In the 33 calls of the benign session of our scripted matrix, 31 need one copy and 2 need two; extra copies arise only when a call's destinations or consumers differ in view or clearance, so these counts reflect our workload rather than deployed sessions. Sequential copies also reduce a
timing channel between copies from 37.9--47.7\,bit/s to at most 0.61\,bit/s~\cite{lampson}.

\para{Memory and end-to-end time.}
A live copy adds 20\,MB PSS (0.98\,MB private) to a 35\,MB germline, and each resident germline about 12\,MB (\cref{fig:eval}(c)). The median LLM-driven episode took 2.04\,s under \sys{} and 1.81--1.97\,s under the baselines.

\section{Related Work}\label{sec:related}

\begin{table}[t]
\centering
\caption{Protection each defence class gives against a signed, functionally correct release with hidden code. $\bullet$~provided; $\circ$~partial or detection only; --~not provided; n/a~outside the class's purpose; $^{\dagger}$~measured here, other cells from the cited designs. Assumptions: \emph{Sign}~\cite{tuf,sigstore,nvidiaskills}, trusted key; \emph{Scan}~\cite{biv,biflow}, readable code; \emph{Honey}~\cite{skillsentry}, decoys reach the hidden code; \emph{Confine}~\cite{agentbound,mcpguard}, correct allowlists; \emph{IFC}~\cite{fides,camel,adams}, replies depend only on the current call; \emph{Attest}~\cite{aclemcp}, workload identity; \emph{Reset}, snapshot restore~\cite{groundhog} (design) or fresh process per call (measured); \sys{}, \asm{1}--\asm{6}, policy labels, declared destinations.}
\label{tab:novelty}
\footnotesize
\setlength{\tabcolsep}{1.2pt}
\begin{tabular}{@{}>{\raggedright\arraybackslash}p{0.33\columnwidth}*{8}{c}@{}}
\toprule
Blocks or provides & Sign & Scan & \shortstack{Hon-\\ey} & \shortstack{Con-\\fine} & IFC & \shortstack{Att-\\est} & \shortstack{Re-\\set} & \shortstack{\textsc{Germ-}\\\textsc{line}} \\
\midrule
Undeclared hosts          & --$^{\dagger}$ & $\circ^{\dagger}$ & $\circ^{\dagger}$ & $\bullet^{\dagger}$ & n/a & -- & -- & $\bullet^{\dagger}$ \\
Unrelated file reads      & --$^{\dagger}$ & $\circ^{\dagger}$ & $\circ^{\dagger}$ & $\bullet^{\dagger}$ & n/a & -- & -- & $\bullet^{\dagger}$ \\
Cross-call carry-over     & --$^{\dagger}$ & $\circ^{\dagger}$ & $\circ^{\dagger}$ & --$^{\dagger}$ & --$^{\dagger}$ & -- & $\circ^{\dagger}$ & $\bullet^{\dagger}$ \\
Within-call mixing        & --$^{\dagger}$ & $\circ^{\dagger}$ & $\circ^{\dagger}$ & --$^{\dagger}$ & --$^{\dagger}$ & -- & --$^{\dagger}$ & $\bullet^{\dagger}$ \\
Dormant logic             & --$^{\dagger}$ & $\circ^{\dagger}$ & --$^{\dagger}$ & --$^{\dagger}$ & --$^{\dagger}$ & -- & $\bullet^{\dagger}$ & $\bullet^{\dagger}$ \\
Stateful-tool labels      & -- & -- & -- & -- & --$^{\dagger}$ & -- & $\circ$ & $\bullet^{\dagger}$ \\
Per-destination views     & n/a & n/a & n/a & n/a & $\circ$ & n/a & n/a & $\bullet^{\dagger}$ \\
Verifiable evidence       & -- & -- & -- & -- & -- & -- & -- & $\bullet^{\dagger}$ \\
\bottomrule
\end{tabular}
\end{table}

\para{Attacks, supply chain, confinement and action control.}
Beyond the attacks of \cref{sec:intro}~\cite{skilltrojan,beyondtoolpoisoning,certeu}, skills deviate from declared capabilities~\cite{biv}, MCP servers contain risky bidirectional flows~\cite{biflow}, and tool poisoning~\cite{owaspmcp} and parser exploits~\cite{cve202564512} add entry points. TUF~\cite{tuf}, in-toto~\cite{intoto} and Sigstore~\cite{sigstore} authenticate who published an artefact and how it was built, which can cover hidden code~\cite{xz,thompson}. AgentBound~\cite{agentbound} and MCPGuard-Dynamic~\cite{mcpguard} build on privilege separation~\cite{privsep,wedge}, capabilities~\cite{capsicum} and sandboxing~\cite{landlock} to bind authority to a process and its destinations. Attestation~\cite{aclemcp}, provenance binding~\cite{decodetime}, trace checks~\cite{vigil}, transactional staging~\cite{cordon}, action validation~\cite{parallax} and action logs~\cite{tracekit,msagt} decide whether an action may happen, not which items it carries.

\para{Information flow, multi-execution and accountability.}
FIDES~\cite{fides} and CaMeL~\cite{camel} label tool replies and enforce flow policies in the planner, and TACIT restricts agent code with capture-checked types~\cite{tacit}; their labels are sound when the reply depends only on the current call, whereas \cref{cor:labels} also joins stored state. Our confinement premise follows Lampson~\cite{lampson}. Secure multi-execution~\cite{sme}, faceted values~\cite{faceted} and multi-execution for language models~\cite{adams} run a program once per security level, DIFC and taint tracking~\cite{asbestos,histar,flume,taintdroid} propagate labels, and TightLip~\cite{tightlip} and N-variant systems~\cite{nvariant} compare replicas. Groundhog restores a FaaS process to a clean snapshot between requests~\cite{groundhog}. PeerReview~\cite{peerreview} and Arbitrum~\cite{arbitrum} make faults provable by replay, as our witnesses do for tool releases, and dampening adapts BGP route-flap damping~\cite{rfc2439}; \cref{sec:bg-novelty} states what germline execution adds to these ideas.

\section{Discussion}\label{sec:discussion}

\subsection{Deployment and composition}\label{sec:discussion-composition}

MCPGuard and AgentBound decide where a tool process may connect~\cite{mcpguard,agentbound}, and germline execution which data each call carries there; each copy can run under the server's kernel-level policy, so the two stack (not measured). The main policy burden is authoring labels and clearances, as in FIDES and CaMeL~\cite{fides,camel}, and declaring destinations in the manifest~\cite{agentbound}. Stateful tools remain compatible when their state passes through the broker's slots (\cref{sec:eval-real}); state kept only in process memory does not carry over, because every copy starts from the data-free germline. Concurrency, throughput, many resident releases and cloud-hosted services are future work.

\subsection{Scope and boundary conditions}\label{sec:scope}

Within the disclosure-only scope of \cref{sec:threat-vectors}, four boundaries apply. \para{Remote tools.}
A tool we do not host cannot be forked or twinned, so it receives prevention through worst-case reply labels, at a utility cost (\cref{sec:eval-labels}), but no witnesses.

\para{Hardware nondeterminism.}
A release that reads the CPU cycle counter (\fn{rdtsc}) still cannot disclose (\cref{thm:sni} does not use \asm{4}), but its divergences stay unverified (\cref{prop:nondet}).

\para{Timing channels.}
Covert timing and microarchitectural channels lie outside \asm{1}; sequential copies reduce the measured channel (\cref{sec:eval-overhead}).

\para{Labels from policy and integrity of effects.}
\Cref{thm:sni} protects data as labelled and bounds what a request carries, not whether it is wanted: a harmful data-free request, such as a delete command, needs action control~\cite{mcpguard,agentbound,cordon,parallax}. A flow-compliant tool that passes data across horizons loses that function and fails closed (\obl{4}, \cref{tab:obligations}).

\section{Conclusion}\label{sec:conclusion}

A signature tells an agent who published a tool, and a confinement policy where it may connect, but neither tells it which data each call carries. Germline execution runs each call in copies forked from a data-free germline, provides sink noninterference for every signed release under the stated assumptions, and ties each verified flow violation to the exact release, at 10--49\,ms per single-copy call.

\section*{Acknowledgments}
\textit{Author contributions.} Aun Yichiet conceived the research problem and the germline-execution mechanism, designed the threat model, the attacks and the experiments, built and ran the testbed, and wrote and edited the manuscript. Mohammed A. Alhartomi reviewed the manuscript and identified inconsistencies that were corrected before submission. Tan Jing Jie provided technical editorial review. Waqas Nadeem cross-checked the generated Trojan designs. Adeb Salh, Shankar Karuppayah and Tan Sin Chee reviewed the conceptual soundness of the work: the threat model, the assumptions and the correctness of the mechanism.

\textit{Use of AI tools.} The authors used AI assistants (Anthropic Claude and OpenAI Codex) as assistive tools for code implementation, experiment execution, data processing and language editing, and a locally served Qwen3.5-27B model generated the attack-variant corpus described in \cref{sec:eval-setup}. The research ideas, the mechanism design, the experimental methodology and the scientific conclusions are the authors' own. The authors reviewed and verified all AI-assisted output and take full responsibility for the content of this paper.

\bibliographystyle{IEEEtran}
\bibliography{refs}

\clearpage
\appendices
\crefalias{section}{appendix}\crefalias{subsection}{appendix}\crefalias{paragraph}{appendix}
\section{Full Proofs}\label{app:proofs}

This appendix proves the statements of \cref{sec:formal}.
Each proof begins with a one-line restatement and a sentence of intuition; the definitions, assumptions and labels are those of \cref{sec:formal-model,sec:formal-assumptions}, and the mechanism is \Plan{} together with \cref{alg:call}.
We use four structural facts, which hold by construction.
(P1)~\Plan{} groups $\Decl_f\cup\Cons_\call$ by view, so all members of a group $g$ share one view, written $\View(g)$.
(P2)~After the split step, every destination host $e$ in a group $g$ satisfies $\clr(e)\subseteq\clr(d)$ for every member $d$ of $g$; since $e$ is itself a member, all destination hosts in $g$ have the same clearance, and it equals the horizon $\hor(g)$.
(P3)~Groups run one at a time in an order in which $\View(g')\subsetneq\View(g)$ implies that $g'$ runs before $g$; the order depends only on the inputs of \Plan{}.
(P4)~There is one transcript and replay semantics, that of \cref{def:responses}: every reply that a copy receives from a host outside its own group is the reply the positional rule gives from that host's record at that moment, and the probe and the control receive the replies the same rule gives from $T(d)$.
In particular, the copy serving $d$ is itself an execution under $\Resp(d)$, because its transcript $T(d)$ records exactly the replies it was served, and replaying $T(d)$ by position to the same requests returns them again.

\subsection{Slot invariant}

\para{Restatement of \cref{lem:slot}.}
Every value in $\slot{\hor}$ derives only from items labelled inside $\hor$, earlier contents of $\slot{\hor}$, public metadata, twin inputs and replies from hosts $e$ with $\clr(e)\subseteq\hor$, and its stored label lies inside $\hor$.
The intuition is that a copy of horizon $\hor$ never holds anything from outside $\hor$, so nothing it writes can contain such data.

\begin{proof}
Call a value \emph{$\hor$-bounded} if it is computed only from the inputs listed in the lemma.
We prove, by induction over the global sequence of broker messages (calls in order, groups in \Plan{} order, messages in order), the stronger claim that every message sent by a copy of horizon $\hor$, whether a request, a state write or a result, is $\hor$-bounded, and that every label it stores lies inside $\hor$.
The lemma follows, because \cref{alg:call} writes $\slot{\hor}$ only on a state write from a copy whose horizon is $\hor$.

Consider a copy of group $g$ with $\hor(g)=\hor$.
At fork its memory is the germline image $\Img$, which depends only on $\rel$ (\asm{2}).
The call message adds the metadata $\meta_\call$, which is public (\asm{5}); the twin inputs $\twin_\call$, which the broker chooses without reading item contents (\cref{def:twin}); the contents of the items in $\View(g)$; and the whole slot $\slot{\hor}$ with its label.
By (P1), each item $x\in\View(g)$ satisfies $\lab(x)\subseteq\clr(d)$ for every member $d$ of $g$, hence $\lab(x)\subseteq\hor$, and the slot contents are $\hor$-bounded, with labels inside $\hor$, by the induction hypothesis.
While it runs, the copy's only input is the broker socket (\asm{1}), and each broker reply takes one of three forms.
(i)~A live reply from a destination host $e$ in $g$: by (P2), $\clr(e)=\hor$.
(ii)~A reply to a request to a declared host $e$ outside $g$, given by the positional rule (P4): the $j$-th recorded reply of $e$ if the digests match and $e$ is admissible for $g$, that is, $\clr(e)\subseteq\hor$, and the placeholder otherwise.
(iii)~A denial or a state acknowledgement: these are constants.
Reaching the deadline gives the copy no input; it ends the copy.
It remains to check that the \emph{outcome} of the positional rule in form (ii) leaks nothing outside $\hor$.
If $e$ is not admissible for $g$, the broker returns the placeholder whatever the requests are, so the outcome is constant; the same holds if $e$'s group has not yet run, because its record is still empty and the order of groups is fixed by (P3).
If $e$ is admissible, the outcome depends on $g$'s own $j$-th request to $e$ and on the record of $e$, that is, the requests delivered to $e$ by an earlier group $g_e$ and their replies.
By the induction hypothesis that record is $\hor(g_e)$-bounded, and by (P2) $\hor(g_e)=\clr(e)\subseteq\hor$, so it is $\hor$-bounded as well; a returned recorded reply carries the label $\clr(e)\subseteq\hor$ (\asm{6}).
The copy is a program whose messages are a function of its initial memory and its input stream, and both are $\hor$-bounded; hence so is every message it sends.
Finally, the label stored with a state write is the join of $g$'s view labels, the label of the slot it received and the clearances of hosts whose replies reached it, each of which lies inside $\hor$ by (P1), the induction hypothesis, (P2) and admissibility.
\end{proof}

\subsection{Sink noninterference}

\para{Restatement of \cref{thm:sni}.}
Under \asm{1}--\asm{3}, \asm{5} and \asm{6}, for every release $\rel$, the request sequence delivered to each declared destination host $d$ satisfies \cref{eq:sni}.
The intuition is that only the copy of $d$'s group reaches $d$, and that copy starts from the data-free image with only $d$'s view, $d$'s slot and public inputs, and later receives only admissible replies or constants.

\begin{proof}
Fix a call $\call$ and $d\in\Decl_f$, and let $g$ be the \Plan{} group that contains $d$.
By (P1), $\View(g)=\View_\call(d)$, and by \cref{def:horizon} the copy of $g$ receives and writes only $\slot{\hor_\call(d)}$.
We proceed in four steps.

\emph{Step 1: only $g$ reaches $d$.}
\Cref{alg:call} delivers a request to $d$ only when it comes from the copy of the group that contains $d$.
A request to $d$ from any other copy is answered by the positional rule and never delivered, and a request to a host outside $\Decl_f$ is denied.
Hence $\Deliv_\call(d)$ consists exactly of the requests that $g$'s copy sends to $d$, in order.

\emph{Step 2: $g$ starts from allowed state.}
At fork the copy's memory is $\Img$.
By \asm{2}, $\Img$ depends only on $\rel$: check \gchk{1} ensures that no second thread exists that could have received data, \gchk{2} ensures that no descriptor exists through which data could arrive later, and \gchk{3} ensures that no other copy can write into the copy's memory.
The call message then adds $\meta_\call$ (public by \asm{5}), the contents of the items in $\View_\call(d)$, the slot $\slot{\hor_\call(d)}$ and $\twin_\call$; items outside $\View_\call(d)$ arrive as names with empty contents.

\emph{Step 3: $g$ receives only allowed inputs while it runs.}
By \asm{1}, the copy's only channel is its broker socket, and by \asm{3} the broker behaves as specified.
By (P4), each reply to a request to a declared host is a live reply of a host in $g$, which is admissible by (P2), or the outcome of the positional rule, which is an admissible recorded reply or the placeholder; together these are exactly the answers of the reply environment $\Resp_\call(d)$ to $g$'s requests.
The remaining broker messages, denials and state acknowledgements, are constants.
By the proof of \cref{lem:slot}, every reply in $\Resp_\call(d)$ and every outcome of the positional rule is $\hor_\call(d)$-bounded; in particular, the records it draws on come from groups $g_e$ with $\View(g_e)\subseteq\View_\call(d)$, because $\lab(x)\subseteq\clr(e)\subseteq\clr(d)$ for every item they hold, so $\Resp_\call(d)$ does not depend on the contents of items outside $\View_\call(d)$.

\emph{Step 4: conclusion.}
A copy is a program whose $j$-th message is a function of its initial memory and the replies to its earlier messages.
By Steps 2 and 3, the initial memory is a function of $\meta_\call$, $\restr{\Items_\call}{\View_\call(d)}$, $\slot{\hor_\call(d)}$ and $\twin_\call$, and the replies are given by $\Resp_\call(d)$.
Induction over the sequence of broker messages shows that the copy's request sequence to $d$ is such a function, and by Step 1 it is the sequence delivered to $d$.
If the copy is killed at the deadline, $d$ receives a prefix of that sequence; which prefix is a matter of timing, which \asm{1} places outside the model.
A deadline can change the inputs of a later copy only through the length of a record: if the copy of a group $g_e$ is killed, the record of each host $e$ in $g_e$ is truncated, and a later copy to which $e$ is admissible receives placeholders at the missing positions.
That record lies inside $\clr(e)\subseteq\hor_\call(d)$ by Step~3, and its length is again timing, so the argument above is unchanged for every combination of deadlines.
If $\rel$ draws on randomness that the broker does not control, the copy is a randomised program, and the same induction shows that the distribution of its output stream is a function of the same five arguments.
\end{proof}

\para{Restatement of \cref{cor:sessions}.}
Across a session, what $d$ receives depends only on items labelled inside $\clr(d)$, from this and earlier calls, and on admissible replies.
The intuition is that the only path from an earlier call to $d$ runs through $d$'s slot, which \cref{lem:slot} bounds.

\begin{proof}
We argue by induction over the calls of the session.
In call $\call$, \cref{thm:sni} restricts $\Deliv_\call(d)$ to the five arguments of \cref{eq:sni}.
The metadata is public and the twin inputs are chosen without reading contents.
Every item of $\View_\call(d)$ has a label inside $\clr(d)$ by \cref{def:view}.
Because $d$ belongs to the group whose horizon is $\hor_\call(d)$, we have $\hor_\call(d)\subseteq\clr(d)$, and by \cref{lem:slot} the slot $\slot{\hor_\call(d)}$ holds only values derived from items, replies and earlier slot contents labelled inside $\hor_\call(d)$, from this and earlier calls.
Finally, every reply in $\Resp_\call(d)$ comes from a host $e$ with $\clr(e)\subseteq\hor_\call(d)\subseteq\clr(d)$ and is treated as data labelled $\clr(e)$ (\asm{6}), and by Step~3 of the proof of \cref{thm:sni} the outcome of the positional rule is itself $\hor_\call(d)$-bounded.
Hence every argument depends only on information labelled inside $\clr(d)$ and on admissible replies, which proves the claim for call $\call$ given the claim for earlier calls.
\end{proof}

\para{Restatement of \cref{cor:labels}.}
The label $\lab(g)$ of \cref{eq:label}, attached to a value returned to a consumer $u$ in group $g$, lies inside $\clr(u)$ and covers every item, slot entry and reply that reached the copy.
The intuition is that each of the three joins names one kind of input, and each kind lies inside $g$'s horizon.

\begin{proof}
\emph{Within clearance.}
By (P1), $\View(g)=\View_\call(u)$, so every item $x\in\View(g)$ satisfies $\lab(x)\subseteq\clr(u)$.
For $e\in E_g$, either $e$ is a destination host of $g$, so $\clr(e)=\hor(g)$ by (P2), or its recorded reply was admissible for $g$, so $\clr(e)\subseteq\clr(d)$ for every member $d$ of $g$; in both cases $\clr(e)\subseteq\hor(g)\subseteq\clr(u)$.
By \cref{lem:slot}, every label stored in $\sigma_g=\slot{\hor(g)}$ lies inside $\hor(g)$, so $\lab(\sigma_g)\subseteq\hor(g)\subseteq\clr(u)$.
All three joins of \cref{eq:label} therefore lie inside $\clr(u)$.

\emph{Covering every input.}
By Steps 2 and 3 of the proof of \cref{thm:sni}, with $u$ in place of $d$, the returned value is a function of the data-free image $\Img$, the public metadata, the twin inputs, the contents of the items of $\View(g)$, the entries of the slot $\sigma_g$ delivered at call entry, and the replies the copy received.
Item labels are joined in the first term.
Each reply other than the placeholder comes from a host in $E_g$ and carries the label $\clr(e)$ (\asm{6}), which the second term joins; the placeholder is a constant, and the fact that the positional rule returned it is the input discussed in the remark below.
For the slot, we show by induction over slot writes that every entry depends only on inputs whose labels lie inside the label stored with it.
When a copy writes an entry, its value depends only on the copy's items, the slot it received and the replies received before the write; the stored label joins exactly the labels of these items, the label of that slot, which covers its entries by the induction hypothesis, and the clearances of those replying hosts.
Hence every entry of $\sigma_g$ is covered by $\lab(\sigma_g)$, which the third term joins.
Because the whole slot is delivered at call entry, this term is needed whenever the slot is non-empty, whether or not the tool reads it, and it is $\emptyset$ for an empty slot.
\end{proof}

\begin{remark}
\Cref{eq:label} is the label the prototype computes (\cref{alg:call}): it joins the labels of the copy's view, the clearance of every host whose reply reached the copy, including admissible recorded replies from hosts outside $g$ (\obl{4}), and the label of the delivered slot.
The label regression test (\cref{tab:regression}) checks the slot term: an HR item stored by one call and recalled by a later call with only public inputs is returned with the label \emph{hr}, so a downstream policy engine refuses to forward it to a public sink, whereas a public-only control keeps the label $\emptyset$ and is forwarded.
Joining the horizon $\hor(g)$ instead of $\lab(\sigma_g)$ for a non-empty slot is also sound but coarser, because it labels a recalled public entry as if it held everything the horizon admits.
The second term also covers the match outcome: the positional rule tells a copy whether its $j$-th request to an admissible host $e$ outside $g$ matched the recorded one, since it receives either the recorded reply or the placeholder, so the prototype joins $\clr(e)$ whenever the copy sends a request to an admissible host with a non-empty record, matched or not.
This bit concerns data inside $\clr(e)\subseteq\hor(g)\subseteq\clr(u)$; the match-bit regression test (\cref{tab:regression}) shows the label of a mismatched copy now includes $\clr(e)$.
\end{remark}

\subsection{Witnesses and detection coverage}

All runs compared below are executions under the reply environment $\Resp(d)$ of \cref{def:responses}: the copy of $d$'s group $g$ by (P4), and the probe and the control by \cref{def:probe}.
In $g$, the items of $W\setminus\View(d)$ arrive as names with empty contents, and their names and labels are public metadata.
The probe and $g$ therefore agree on the metadata, on the contents of $\View(d)$, on the slot snapshot $\slot{\hor(d)}$, on $\twin$ and on $\Resp(d)$, and differ only in the contents of the items in $W\setminus\View(d)$; the control agrees with $g$ on all of its inputs.

\para{Restatement of \cref{thm:witness}.}
For a release that is deterministic within the replay model, (a)~an item witness implies a flow violation toward $d$, (b)~a flow-compliant release never yields an item witness, and (c)~in a multi-copy call, every difference between the delivered sequence and the request sequence to $d$ of a comparing copy whose view strictly contains $\View(d)$ and that finishes within its deadline raises a trigger that probe replay classifies.
The intuition is that replay equalises every argument $d$ may depend on, so any remaining difference must come from items $d$ may not see.

\begin{proof}
(a)~Let $\rel$ be deterministic within the replay model, and suppose an item witness for $(\rel,d)$ exists, so the probe's request sequence to $d$ differs from the delivered one.
By the observation above, the probe and $g$ agree on all five arguments of \cref{eq:sni}, including the reply environment through which each of them is answered, and differ only in contents of items outside $\View(d)$.
Because $\rel$ is deterministic within the replay model, each execution's request sequence is a function of its inputs and its reply environment, so the two executions agree on the five arguments yet send different request sequences to $d$, which is a flow violation by \cref{def:sni}.
The $N$ repeated tries and the control are not needed for this implication when $\rel$ is deterministic; they make the verdict robust in practice, because a release that draws on a source outside \asm{4} can fail the control and is then never reported as an item witness.

(b)~If $\rel$ is flow-compliant toward $d$, its request sequence to $d$ is a function of the five arguments of \cref{eq:sni}.
The probe and $g$ agree on all five, so they send identical sequences to $d$ on every try, and no item witness is produced.
The argument uses the slot snapshot $\slot{\hor(d)}$ and the reply environment $\Resp(d)$ in both runs, so it applies to stateful releases as well as stateless ones.

(c)~Let $g'$ be a copy with $\View(g')\supsetneq\View(d)$ that finishes within its deadline; by (P3) it runs after $g$, so when $g'$ runs the record of $d$ holds every request delivered to $d$: the complete sequence if $g$ finished, and otherwise the prefix delivered before $g$ was killed at the deadline.
For the $j$-th request of $g'$ to $d$, the broker compares its digest with that of the $j$-th request delivered to $d$, and records a trigger at the first position where they differ or where the record has no $j$-th entry; when $g'$ ends, it records a trigger if $g'$ sent fewer requests to $d$ than were delivered, including none.
A difference in content, count or order, or a request absent from either sequence, produces a first position at which the two sequences differ, or a shorter sequence from $g'$, so each such difference raises a trigger.
Probe replay then runs the probe and the control $N$ times each.
If the control ever differs from the delivered sequence, the trigger is classified as nondeterminism outside \asm{4} (\cref{prop:nondet}); otherwise, if the probe differs on all $N$ tries, it is an item witness, and if it never differs, it is a state or response divergence (\cref{prop:state}); a probe that differs on some tries only is also classified as nondeterminism.
For a release that is deterministic within the replay model and whose copy $g$ finished, the control never differs and every probe try gives the same result, so the trigger is an item witness, which is a flow violation by (a), or a state or response divergence.
If $g$ was killed, the record is a prefix whose length is a matter of timing (\asm{1}); whenever the control does not reproduce that prefix, the trigger is classified as nondeterminism and not as an item witness.
\end{proof}

\begin{proposition}[State and response divergences]\label{prop:state}
A trigger whose probe matches the delivered sequence shows that the comparing copy's request depended on its own slot or on replies that $d$'s copy was not served; it is prevented by \cref{thm:sni} but not verified.
\end{proposition}

Dampening gives such an event the weight of an unverified event, 250 at the operating point (\cref{tab:replay}).

\begin{proof}
The comparing copy and the probe hold the same items $W$, metadata and twin inputs, and differ only in the slot ($\slot{\hor(g')}$ against $\slot{\hor(d)}$) and in the reply environment: the comparing copy is answered live by its own hosts and by the positional rule with its own admissibility, the probe by $\Resp(d)$.
If the comparing copy's sequence to $d$ differs from the delivered one while the probe's equals it, then, by determinism, the difference is caused by the slot or the replies, not by the contents of $W\setminus\View(d)$.
The comparing copy's requests are never delivered to $d$ (Step~1 of the proof of \cref{thm:sni}), so nothing is disclosed.
A flow-compliant tool that keeps a call counter per horizon produces exactly this pattern, because counters in different slots differ; hence the event cannot be attributed to a flow violation and receives the unverified weight (\cref{tab:replay}).
\end{proof}

\begin{proposition}[Nondeterministic releases]\label{prop:nondet}
If $\rel$ draws on sources outside \asm{4}, the control can fail and replay cannot confirm a flow violation; a release that declared determinism in its signed metadata and fails the control contradicts a declaration its signature attributes to the publisher.
\end{proposition}

\begin{proof}
If the control, run on $g$'s exact inputs under $\Resp(d)$, produces a sequence different from the delivered one, then the output of $\rel$ is not a function of the inputs that the broker controls, so $\rel$ draws on a source outside \asm{4}.
The argument of \cref{thm:witness}(a) needs determinism to attribute a difference between the probe and $g$ to item contents, so it no longer applies, and no item witness is reported.
A release that declared determinism in its signed metadata and fails the control contradicts a statement that its signature binds to the publisher key; by \asm{3}, signature verification is correct, so the broken declaration is attributable to the publisher.
\end{proof}

\subsection{Copies per call}

The mechanism class is the one stated in \cref{thm:copies}.
A mechanism in the class treats the release as a black box: it learns the release's behaviour only by running it, and every request it delivers and every value it returns is produced by some run of the release on some item set $U\subseteq\Items_\call$, in which the items outside $U$ arrive as names with empty contents.
A data-independent fork rule (\cref{def:forkrule}) fixes these runs, and the output rule that assigns to each member of $\Decl_f\cup\Cons$ the run whose output it receives, before any run starts and without reading the release's code; two releases with the same manifest therefore receive the same runs and the same output rule.
For consumers, we read the requirement of \cref{thm:sni} in its natural form for return values: a consumer receives nothing derived from items outside its view.

\para{Restatement of \cref{thm:copies}.}
(a)~A mechanism in this class that prevents disclosure for every release and is transparent needs a run on each distinct view among $\Decl_f\cup\Cons$ and, within a view, one per distinct host clearance plus one for the consumers whose clearance contains no host clearance of that view; (b)~\Plan{} creates exactly these runs.
The intuition is that a flow-compliant release can encode its whole view into each output, so every distinct view, and every clearance that changes which replies a run may see, needs its own run.

\begin{proof}
\emph{(a), views.}
Fix a call and a member $d$ of $\Decl_f\cup\Cons$, and consider two releases with the same manifest.
Release $\rel_1$ sends to every declared destination host, and returns to every consumer, an injective encoding of the contents of all items it holds.
Release $\rel_2$ sends to each declared host $e$, and returns to each consumer $e$, an injective encoding of the contents of the items $x$ with $\lab(x)\subseteq\clr(e)$; it can compute this set because labels and clearances are public.
Release $\rel_2$ is flow-compliant, whereas $\rel_1$ is not.
The output rule assigns to $d$ one run, on some item set $U$.
Because the runs and the output rule are fixed by a data-independent rule from $f$, $\meta$, labels, clearances, $\Decl_f$ and $\Cons$, which are the same for $\rel_1$ and $\rel_2$, the rule assigns to $d$ a run on the same set $U$ under both releases.
This step uses the data independence of the output rule and not only of the runs: a mechanism that fixed its runs in advance but chose which run's output to deliver after observing the outputs could choose differently for $\rel_1$ and $\rel_2$.
If $U$ contains an item outside $\View(d)$, then under $\rel_1$ what $d$ receives depends on that item's contents, which contradicts \cref{thm:sni}.
If $U$ misses an item $y\in\View(d)$, then under $\rel_2$ what $d$ receives does not depend on $y$'s contents, because $y$ arrives empty, while transparency requires the injective encoding of $\restr{\Items}{\View(d)}$, which does; for some contents of $y$ the two differ.
Hence $U=\View(d)$, and since the rule is fixed before any run, the mechanism needs a run on each distinct view among $\Decl_f\cup\Cons$.

\emph{(a), groups.}
Let $e_1$ and $e_2$ be destination hosts with the same view and $\clr(e_1)\not\subseteq\clr(e_2)$.
Consider a flow-compliant release that first sends a request to $e_1$, then sends to $e_2$ an injective encoding of the reply it received from $e_1$, and then sends to $e_1$ a second request that also encodes that reply.
The reply of $e_1$ is not admissible for $e_2$, so by \cref{def:transparency} the transparent sequence for $e_2$ is computed with the placeholder in its place, while the transparent sequence for $e_1$ is computed with $e_1$'s real reply.
A single run receives either the real reply or the placeholder.
With the real reply, its request to $e_2$ differs from the transparent one whenever the reply differs from the placeholder, and it also carries $e_1$'s data to $e_2$; with the placeholder, its second request to $e_1$ differs from the transparent one.
Hence $e_1$ and $e_2$ need separate runs.
The same argument, with a return value in place of the request to $e_2$, separates a consumer from any destination host whose clearance is not inside the consumer's.

\emph{(b).}
Within each view, \Plan{} partitions the destination hosts by clearance, attaches each consumer to a part whose clearance lies inside its own, and places the remaining consumers of that view together in one part.
Two hosts with different clearances violate inclusion in at least one direction, so the groups argument of (a) forces one run per distinct clearance among the hosts of a view.
A remaining consumer has no host in its view whose clearance lies inside its own, so (a) forces it to run apart from every host part.
Remaining consumers can share a run, because consumers return no replies and every host reply is inadmissible for all of them.
Every part \Plan{} creates is therefore forced, and \Plan{} creates no others, so its number of copies equals the lower bound of (a) within this class.
\end{proof}

\para{Restatement of \cref{lem:existence}.}
Under lazy splitting, whether $d$ receives any request reveals one bit of uncleared data per call, invisibly to the tripwire.
The intuition is that the existence of $d$'s copy, and hence of any request to $d$, is decided by a copy that holds data $d$ may not see.

\begin{proof}
Under lazy splitting, the copy with view $\View(d)$ is forked only when the full-view copy contacts $d$.
Consider a release that, in the full-view copy, contacts $d$ if and only if a chosen bit $b$ of an item outside $\View(d)$ equals 1, and whose request to $d$ does not depend on any uncleared content.
Then $d$ receives a request exactly when $b=1$, so the presence of a request reveals $b$: one bit per call and destination host.
When $b=1$, the delivered request comes from the copy with view $\View(d)$ and, by construction, equals the full-view copy's request; when $b=0$, no copy for $d$ exists and nothing is compared.
In neither case does the tripwire record a trigger.
Under declared-destination splitting, the copy for $d$ is forked whenever $d\in\Decl_f$, independently of item contents, and the bit disappears.
In the existence-bit test, when the copy that cannot see the secret falls back to sending a request, which is the release constructed above, lazy splitting recovers 64 of 64 secret bits (0.942 bits per call); declared-destination splitting recovers no information under either fallback (0.000 bits per call; accuracy at the constant-guess baseline); with declared-destination splitting the sequence comparison also raises a trigger on every secret-dependent attempt, 23 of 23 when the release sends only if the bit is 1 and 41 of 41 when it sends only if the bit is 0 (\cref{sec:eval-adaptive}).
\end{proof}

\para{Cost of a call.}
A call with $n\le|\Decl_f\cup\Cons|$ groups and $q$ requests costs $n$ forks, $n$ tool runs and $n$ germline checks, and $O(q)$ digest comparisons; verifying a trigger takes $2N$ replays (probe and control) off the critical path.

\section{Additional Results}\label{app:tables}

This appendix maps every experiment to its research question and result file, gives the complete result matrices
behind the summary tables of \cref{sec:eval}, reports the tests that connect the proofs to the prototype, and
analyses the dampening policy. Experiment identifiers (E1--E13) appear only in this appendix and in
\cref{app:openscience}.

\subsection{Experiment map}\label{app:eid}

\Cref{tab:eid-map} links each experiment to the research question it answers, the subsection that reports it, its
result file, the platform it ran on and the code it measured. Every experiment marked M ran on the final
implementation described in \cref{sec:design,sec:impl}, with slot labels, the match bit, sequence comparison, the
absolute per-copy deadline and the broker without redirects, and with determinism read from the signed manifest.

\begin{table*}[tbp]
  \centering
  \caption{Experiment map. Result files are in \fn{results\_cloud/} (C) or \fn{results\_server/evidence\_freeze/}
    (S) of the artifact. Platforms as in \cref{sec:eval-setup}: cloud container (Landlock ABI~7) or server
    (ABI~4). Code: \textbf{M}, the final implementation of \cref{sec:design,sec:impl}; \textbf{I}, independent of the enforcement code (detection baselines, timing channel);
    \textbf{G}, MCPGuard's configurations only, without \sys{} code.}
  \label{tab:eid-map}
  \footnotesize
  \begin{tabular}{@{}llp{0.13\textwidth}p{0.43\textwidth}ll@{}}
    \toprule
    ID & RQ & Reported in & Result file(s) & Platform & Code \\
    \midrule
    E1    & RQ2, RQ3 & \cref{sec:eval-matrix,sec:eval-utility} & C: \fn{e1\_final.json} & cloud & M \\
    E2    & RQ4 & \cref{sec:eval-witness} & C: \fn{e2\_detection.json} & cloud & I \\
    E3    & RQ5 & \cref{sec:eval-overhead} & S: \fn{E3\_final\_pass1.json}, \fn{\_pass2}, \fn{\_pass3} & server & M \\
    E4    & RQ5 & \cref{sec:eval-overhead} & C: \fn{e4\_covert.json} & cloud & I \\
    E5    & RQ5 & \cref{sec:eval-overhead} & S: \fn{E5\_final.json} & server & M \\
    E6    & RQ3, RQ5 & \cref{sec:eval-llm,sec:eval-overhead} & S: \fn{E6\_final\_3rep.json} & server (GPU) & M \\
    E7    & RQ3, RQ5 & \cref{sec:eval-real,sec:eval-overhead} & C: \fn{e7\_real\_mcp.json}, \fn{e7\_more.json} & cloud & M \\
    E8-A  & RQ1 & \cref{sec:eval-mcpguard} & S: \fn{E8A\_refresh\_20261103T140804.json} & server & G \\
    E8-C  & RQ1 & \cref{sec:eval-mcpguard} & S: \fn{E8C\_final.json} & server & M \\
    E9    & RQ4 & \cref{sec:eval-adaptive} & C: \fn{e9\_existence\_bit.json}, \fn{e9\_deniability.json} & cloud & M \\
    E10   & RQ2 & \cref{sec:eval-labels} & C: \fn{e10\_remote\_tool.json}; CaMeL check script and log & cloud & M \\
    E11   & RQ4 & \cref{sec:eval-witness} & C: \fn{e11\_witness.json} & cloud & M \\
    E12   & RQ4 & \cref{sec:eval-dampening}, \cref{app:dampening} & C: \fn{e12\_montecarlo.json}, \fn{e12\_dampening.json}, \fn{e12\_sweep.json}; \fn{e12\_montecarlo\_h100.json} ($\hl=100$) & cloud & M \\
    E13   & RQ2 & \cref{sec:eval-scale} & S: \fn{E13\_final\_3trial.json} & server & M \\
    Twin  & RQ4 & \cref{sec:eval-witness} & C: \fn{e\_twin\_falsealarm.json} & cloud & M \\
    \obl{1}--\obl{5} & RQ4 & \cref{sec:eval-witness}, \cref{tab:obligations} & C: \fn{obligations.json}; S: \fn{OBLIG\_final.json} & both & M \\
    Regression & RQ2--RQ4 & \cref{sec:eval-labels,sec:eval-real,sec:eval-witness}, \cref{tab:regression} & C: \fn{review\_tests.json} & cloud & M \\
    \bottomrule
  \end{tabular}
\end{table*}

\subsection{Security results}

\para{Generation protocol for the generated variants (E13).}\label{app:e13}
We generated the variants ourselves. They target the \fn{read\_note} tool of MCPGuard's signed note server, at the same commit \fn{825fdc0a} as the MCPGuard build of \cref{sec:eval-setup}.
A pipeline built with an LLM coding agent (Codex) prompts a local Qwen3.5-27B model on vLLM~\cite{vllm} (temperature 0.8, seed $13000+i$ for attempt $i$, at most 1800 tokens, reasoning off).
Each prompt asks for one pure Python function \fn{candidate(data, state, phase, triggered)} that returns the effects prescribed for its family (for example, return the stored data on the second phase for carry-over), forbids imports, files, sockets and process calls, and asks for a control and data flow distinct from the eight most recent accepted structures; a static checker enforces these rules, and fixed adapters around the unchanged \fn{read\_note} handler turn the returned effects into requests.
Of 189 attempts, 97 were rejected at generation and 92 were packaged as distinct programs (unique normalised syntax trees) in five families: direct (20), allowed-request (20), dormant (20), triggered (18) and carry-over (14).
We adapted each package to our runner, signed it, and admitted it only if its output matches the clean tool and it discloses a canary with no defence. The adaptation rewrites the packaging (manifest fields, module layout and call scenario) and connects the generated canaries to our oracle; the payloads already sent their requests through \fn{urllib}, which the broker relays, and the admission check confirms that every admitted variant still discloses without a defence. Variants are distinct as normalised syntax trees; we did not measure their semantic similarity within a family.
All 92 are functional, and 91 are admitted: one triggered variant did not disclose without a defence and was dropped.
We tagged each admitted case by where it sends data: 71 to an attacker-owned host that no policy lists, and 20 to a host the policy allows.

\para{MCPGuard in its own runner (E8-A).}
\Cref{tab:e8a-full} reports all seven MCPGuard configurations on the six-call session of \cref{sec:eval-setup}, with three trials each (105 runs).
The positive control behaves as intended: \atk{direct}, a raw-socket transfer to an unlisted host, discloses in every trial under the three configurations without MCPGuard's eBPF programs and in no trial under the four configurations that load them.
\atk{pool} and \atk{smuggle}, which place carry-over and in-manifest data inside the authorised request to the OCR destination host, disclose in every trial under all seven configurations, including \mcg{C-full-percall}.
\atk{launder\_text} does not disclose under any configuration in this runner, including \mcg{C0}, which enforces nothing, so the zero does not reflect enforcement; the variant is therefore measured under our runtimes in E8-C, E1 and E6.
The benign release discloses nothing and completes its session under every configuration.
These rows agree with the single-harness comparison of \cref{tab:mcpguard}.

\begin{table}[tbp]
  \centering
  \caption{E8-A: MCPGuard-Dynamic, all seven configurations, in MCPGuard's own runner. Cells give trials that disclose out of 3, with canaries disclosed per trial in parentheses. Every disclosure by \atk{pool} and \atk{smuggle} reaches the authorised OCR destination host; every disclosure by \atk{direct} reaches the attacker-owned host. The last four configurations load MCPGuard's four eBPF programs.}
  \label{tab:e8a-full}
  \footnotesize
  \setlength{\tabcolsep}{3.5pt}
  \begin{tabular}{@{}lccccc@{}}
    \toprule
    Configuration & \atk{benign} & \atk{direct} & \atk{pool} & \atk{smuggle} & \atk{launder\_text} \\
    \midrule
    \mcg{C0}             & 0/3 & 3/3 (3) & 3/3 (2) & 3/3 (3) & 0/3 \\
    \mcg{C-AB}           & 0/3 & 3/3 (3) & 3/3 (2) & 3/3 (3) & 0/3 \\
    \mcg{C-app}          & 0/3 & 3/3 (3) & 3/3 (2) & 3/3 (3) & 0/3 \\
    \midrule
    \mcg{C-ebpf}         & 0/3 & 0/3     & 3/3 (2) & 3/3 (3) & 0/3 \\
    \mcg{C-full}         & 0/3 & 0/3     & 3/3 (2) & 3/3 (3) & 0/3 \\
    \mcg{C-AB+ebpf}      & 0/3 & 0/3     & 3/3 (2) & 3/3 (3) & 0/3 \\
    \mcg{C-full-percall} & 0/3 & 0/3     & 3/3 (2) & 3/3 (3) & 0/3 \\
    \bottomrule
  \end{tabular}
\end{table}

\para{Security matrix per variant (E1).}
\Cref{tab:e1-variant} breaks the 153 attack runs per runtime of \cref{tab:matrix} down by attack variant, and \cref{fig:heatmap} in \cref{sec:eval-matrix} shows the same counts as disclosure rates.
The breakdown shows which object each defence binds authority to.
Confinement in \rtConf{} stops every variant that needs an unlisted destination host (\atk{direct}, \atk{filesteal}) and every attack on initialisation (\atk{init\_read}, \atk{init\_thread}, \atk{hijack}), but not the remaining ten variants: those that travel inside authorised requests (\atk{pool}, \atk{smuggle}, \atk{launder}, \atk{launder\_text}), those that wait or react to content (\atk{dormant}, \atk{trigger}), the adaptive variants \atk{evasive} and \atk{capsule}, the dependency bump \atk{dep}, and the cross-tool composition \atk{xtool}.
A fresh process per call in \rtFresh{} removes most carry-over but still discloses in 3 of 15 runs each for \atk{evasive}, \atk{launder}, \atk{pool} and \atk{smuggle}, and for \atk{xtool} in every run, because one call still mixes differently labelled items.
The ablation \rtNoSplit{} closes the cross-tool case but still discloses through the within-call variants \atk{evasive}, \atk{launder} and \atk{pool}, which is the gap that declared-destination splitting closes.
\rtGerm{} discloses in none of the 153 runs and records 12 triggers along the way, which feed the witness evaluation of \cref{sec:eval-witness}.
Prevention costs no utility in this matrix: the benign workload completes 99 of 99 steps correctly under every runtime, and the benign release discloses in 0 of 15 runs under \rtGerm{}.
The counts in \cref{tab:matrix,tab:e1-variant} include every detected disclosure in an attack run, including disclosures that the agent makes itself.
Under \rtSig{}, which has no output labels, the agent forwards protected replies itself: in 3 of the 15 benign-release runs, and in all 6 attack runs of \atk{hijack} and \atk{init\_thread} (marked $^{\dagger}$), in which the tool itself discloses nothing.
By the attribution rule of \cref{sec:eval-setup}, these 6 runs are not attributable to the tool, so the tool discloses in 108 of the 153 \rtSig{} runs, as under \rtIFC{}; the gap between 114 and 108 is the agent's own forwarding.
The vector column follows the mapping of \cref{sec:threat-vectors}: the eleven \V{1} variants contribute 105 runs, \atk{trigger} 15 runs to \V{1}+\V{2}, \atk{xtool} 3 runs to \V{1}+\V{3}, and the two laundering variants 30 runs to \V{1}+\V{4}.

\begin{table}[tbp]
  \centering
  \caption{E1 per attack variant: attack runs with a detected disclosure, out of the $n$ runs in the last column (3 trials each). Runtimes as in \cref{sec:eval-setup}; NoSplit is \rtNoSplit{}. $^{\dagger}$The agent forwards protected replies itself; the tool discloses nothing in these runs, so 108 of the 114 \rtSig{} disclosures are attributable to the tool.}
  \label{tab:e1-variant}
  \footnotesize
  \setlength{\tabcolsep}{3pt}
  \begin{tabular}{@{}llccccccc@{}}
    \toprule
    Variant & Vec. & \shortstack{Sig+\\Pin} & \shortstack{Sig+\\IFC} & \shortstack{Conf.\\+IFC} & \shortstack{Fresh\\+IFC} & \shortstack{No-\\Split} & \sys{} & $n$ \\
    \midrule
    \atk{capsule}      & \V{1}    & 3  & 3  & 3 & 0 & 0 & 0 & 3  \\
    \atk{dep}          & \V{1}    & 3  & 3  & 3 & 0 & 0 & 0 & 3  \\
    \atk{direct}       & \V{1}    & 12 & 12 & 0 & 0 & 0 & 0 & 15 \\
    \atk{dormant}      & \V{1}    & 15 & 15 & 6 & 0 & 0 & 0 & 15 \\
    \atk{evasive}      & \V{1}    & 15 & 15 & 9 & 3 & 3 & 0 & 15 \\
    \atk{filesteal}    & \V{1}    & 15 & 15 & 0 & 0 & 0 & 0 & 15 \\
    \atk{hijack}       & \V{1}    & 3$^{\dagger}$  & 0  & 0 & 0 & 0 & 0 & 3  \\
    \atk{init\_read}   & \V{1}    & 3  & 3  & 0 & 0 & 0 & 0 & 3  \\
    \atk{init\_thread} & \V{1}    & 3$^{\dagger}$  & 0  & 0 & 0 & 0 & 0 & 3  \\
    \atk{pool}         & \V{1}    & 6  & 6  & 6 & 3 & 3 & 0 & 15 \\
    \atk{smuggle}      & \V{1}    & 9  & 9  & 3 & 3 & 0 & 0 & 15 \\
    \atk{trigger}      & \V{1}+\V{2} & 12 & 12 & 6 & 0 & 0 & 0 & 15 \\
    \atk{xtool}        & \V{1}+\V{3} & 3  & 3  & 3 & 3 & 0 & 0 & 3  \\
    \atk{launder}      & \V{1}+\V{4} & 9  & 9  & 9 & 3 & 3 & 0 & 15 \\
    \atk{launder\_text}& \V{1}+\V{4} & 3  & 3  & 3 & 0 & 0 & 0 & 15 \\
    \midrule
    Total              &          & 114 & 108 & 51 & 15 & 9 & 0 & 153 \\
    \bottomrule
  \end{tabular}
\end{table}

\para{Remote-tool labels and CaMeL labelling check (E10).}
For a flow-compliant remote tool that mixes sensitivities, the worst-case reply labels of \cref{sec:eval-labels} complete 0 public posts, with 3 attempts blocked, where per-call labels complete 2, with 1 blocked; this is the utility cost noted in \cref{sec:scope}.
CaMeL labels a tool's return value from the tool's identity and that call's arguments and keeps no record of earlier calls. In the check of \cref{sec:eval-labels}, run without a language model on CaMeL's reference interpreter at commit \fn{f083b6b}, a stateful tool that the deployer declares side-effect-free returns, in its second reply, a private email body received in its first call; the reply is labelled public and passes the \fn{send\_email} policy to an untrusted recipient in both NORMAL and STRICT modes, while sending the private body directly is denied. CaMeL's default policy denies the undeclared tool's first call. For FIDES we rely on its paper's description of per-call labelling~\cite{fides}.

\subsection{Implementation correspondence}

\para{Boundary details.}
\Cref{tab:boundary} summarises the enforcement boundary. In more detail, the initialisation allowlist covers the interpreter, libraries, CA and time-zone data, the release and the runtime, with no writable path; the broker's environment is inherited without the \fn{http\_proxy} variables. Broker messages are \fn{effect}, \fn{state\_put} and \fn{result}, and requests to undeclared hosts are denied. Each release has its own broker object, a new clearance gets a new slot, and calls run sequentially; a deployment shared by several users must key slots by user as well. Host aliases are distinct destinations. Broker requests time out after 10\,s or at the copy's deadline, whichever comes first, and redirects are returned to the copy unfollowed. The deadline bounds a copy's wall-clock time, and with it its CPU time; there is no memory quota. Absence of a request is compared, its timing is not.

\para{Obligations.}
The proofs of \cref{app:proofs} rely on five implementation properties, the obligations \obl{1}--\obl{5}.
\Cref{tab:obligations} lists for each obligation what could go wrong without it, the mechanism that discharges it, and a property test on a flow-compliant fixture that exercises it directly.
Each test runs with the mechanism switched off and on, on both platforms (Landlock ABI~7 in the cloud container and ABI~4 on the server), with identical results.

\begin{table*}[tbp]
  \centering
  \caption{Obligations \obl{1}--\obl{5}: what each prevents, its implementation, and its property test with the mechanism switched off and on. Results are identical on Landlock ABI~7 (cloud) and ABI~4 (server).}
  \label{tab:obligations}
  \footnotesize
  \begin{tabular}{@{}lp{0.20\textwidth}p{0.29\textwidth}p{0.17\textwidth}p{0.22\textwidth}@{}}
    \toprule
    & Without the mechanism & Implementation & Property test & Switch off $\rightarrow$ on \\
    \midrule
    \obl{1} & The germline could hold pipes, sockets or other descriptors created during initialisation, or inherited standard streams. &
      Before every fork, check \gchk{2}: only the control socket, \fn{/dev/null} and read-only regular files; standard streams bound to \fn{/dev/null}. &
      Release whose initialisation opens a pipe. & Release starts $\rightarrow$ release refused (\gchk{2}). \\
    \obl{2} & The germline could hold an anonymous shared writable mapping. &
      Before every fork, check \gchk{3}: no shared writable mapping. &
      Release whose initialisation maps shared memory. & Release starts $\rightarrow$ release refused (\gchk{3}). \\
    \obl{3} & A copy's descendants could outlive it. &
      Each copy runs as PID~1 of its own PID namespace, so its descendants terminate with it. &
      Tool that leaves a sleeping child process. & 1 live process after the call $\rightarrow$ 0. \\
    \obl{4} & Recorded replies could reach larger views regardless of clearance, and \Plan{} would not split by clearance. &
      Admissibility test on replies; clearance split in \Plan{}; output labels include the clearances of replying hosts; slots keyed by horizon (\cref{def:horizon}). &
      Consumer cleared for \textit{finance}; reply from a host cleared for \textit{projx}. & 1 copy, consumer receives the reply $\rightarrow$ 2 copies, consumer receives the placeholder. \\
    \obl{5} & Verification would replay each copy with its own slot and without the reply transcript. &
      Witnesses record the reply transcript; \fn{verify()} runs the probe replay (\cref{def:probe}) and the control (\cref{def:control}) and classifies the outcome. &
      Flow-compliant per-horizon call counter. & 1 trigger, confirmed as a flow violation (false positive) $\rightarrow$ probe classifies it as a state divergence (weight 250). \\
    \bottomrule
  \end{tabular}
\end{table*}

\para{Enforcement of the assumptions.}
\Cref{tab:assumptions} maps each assumption of \cref{sec:formal-assumptions} to the mechanism that enforces it in the prototype and to the test that checks it.
\asm{3} is the trusted base and is assumed rather than enforced.
\asm{1} does not depend on the Landlock ABI: Landlock ABIs below~6 do not scope signals and abstract sockets, but each copy runs as PID~1 of its own PID namespace, which terminates every descendant with the copy on all ABIs.
For \asm{2}, check \gchk{1} requires numeric libraries to run single-threaded in the germline, which the germline environment configures (\cref{sec:eval-setup}).

\begin{table*}[tbp]
  \centering
  \caption{Assumptions \asm{1}--\asm{6}: how the prototype enforces each one and how it is tested.}
  \label{tab:assumptions}
  \footnotesize
  \begin{tabular}{@{}p{0.13\textwidth}p{0.42\textwidth}p{0.38\textwidth}@{}}
    \toprule
    Assumption & Enforced by & Prototype status and test \\
    \midrule
    \asm{1} Confinement & Fresh network and IPC namespaces per copy; each copy is PID~1 of its own PID namespace; Landlock read-only allowlist (system paths, package, \fn{/proc/self}) with no writable path; on Landlock ABI~$\ge$6, signals and abstract sockets scoped; copies run one at a time, smallest view first; a copy that exceeds its absolute deadline on the broker's clock is killed with its descendants, its consumers receive an error, and the other copies run as planned. & Enforced (\obl{3} and the deadline). A lingering child process: 1 live process with the PID namespace switched off, 0 with it on, on ABI~4 and ABI~7. A copy that never finishes, one that keeps sending messages and one waiting on a slow destination host are each killed at the deadline, and the next call succeeds (\cref{tab:regression}). Sequential copies reduce the parallel timing channel measured in E4 (37.9--47.7\,bit/s in parallel, at most 0.61\,bit/s sequentially; \cref{sec:scope}). \\
    \asm{2} Data-free germline & The broker checks the germline through \fn{/proc} before every fork (\gchk{1} one thread, \gchk{2} descriptors, \gchk{3} shared mappings) and refuses the release if a check fails; numeric libraries pinned to one thread. & Enforced (\obl{1}, \obl{2}). Releases whose initialisation opens a pipe or maps shared memory are refused, on ABI~4 and ABI~7. \\
    \asm{3} Trusted base & Assumption: kernel, broker, label and clearance policy, signature verification. The broker does not follow redirects, so requests go only to the named destination. & Assumed; redirect test: a 302 and a body-preserving 307 reach the attacker host with redirect-following on (1 request each, 21 item bytes for 307) and 0 requests with it off (\cref{tab:regression}). \\
    \asm{4} Twin scope & The tool host freezes wall-clock time and \fn{datetime}, replaces the monotonic clock and performance counters with a deterministic counter, and seeds \fn{random}, \fn{os.urandom} and \fn{numpy.random} per call. & Enforced as stated; the CPU cycle counter (\fn{rdtsc}) lies outside the twin scope (\cref{sec:scope}); prevention does not depend on \asm{4}. Twin test: 20/20 false witnesses without twin execution, 0/20 with it. E9: secret-dependent noise drawn from \fn{os.urandom} or the performance counter yields 20/20 verified witnesses; noise behind a random coin raises none, because both twins draw the same coin. \\
    \asm{5} Public metadata & All copies receive the same metadata; items outside a copy's view arrive as a name with empty contents. & Holds; the experiments pass only paths and flags as arguments, and secret content travels as labelled items. \\
    \asm{6} Responses & The broker returns a destination host's reply to a copy only if it is admissible, by position in that host's recorded sequence, and a placeholder otherwise; \Plan{} splits hosts of different clearance. & Enforced (\obl{4}). A \textit{finance}-cleared consumer receives a \textit{projx}-cleared host's reply with the clearance split switched off and the placeholder with it on. \\
    \bottomrule
  \end{tabular}
\end{table*}

\para{Evidence and regression tests.}
\Cref{tab:evidence} summarises the evidence tests of \cref{sec:eval-witness}, and \cref{tab:regression} the tests for slot labels, sequence comparison, the per-copy deadline, redirects and the two workflows of \cref{sec:eval-real}.
In both tables, each row compares \sys{} with a reference: the mechanism with the named component switched off, the named ablation, or, for the workflows, the unprotected server.

\para{Replay outcomes.}
\Cref{tab:replay} maps each outcome of probe replay (\cref{sec:design-witness}) to the event the prototype records and to its dampening weight. A verified witness carries the full weight 2000 only for a release that declared determinism, because for any other release the control is a statistical check.

\begin{table}[tbp]
  \centering
  \caption{Replay outcomes, the resulting event (as named by the prototype), and its dampening weight.}
  \label{tab:replay}
  \footnotesize
  \setlength{\tabcolsep}{3pt}
  \begin{tabular}{@{}>{\raggedright\arraybackslash}p{2.05cm}>{\raggedright\arraybackslash}p{2.25cm}%
      >{\raggedright\arraybackslash}p{1.95cm}>{\raggedright\arraybackslash}p{0.95cm}@{}}
    \toprule
    Replay outcome & Meaning & Event & Weight \\
    \midrule
    Probe differs on all $N$ tries; control never differs
      & Item witness: requests depend on uncleared items (\cref{thm:witness})
      & \fn{verified} if determinism declared, else \fn{verified\_\allowbreak undeclared}
      & 2000 / 250 \\
    Probe matches; trigger came from slot or replies
      & State or response divergence (\cref{prop:state})
      & \fn{unverified} & 250 \\
    Control differs; determinism declared
      & Signed determinism declaration not kept (\cref{prop:nondet})
      & \fn{declared\_\allowbreak broken} & 2000 \\
    Control differs; determinism not declared
      & Nondeterminism beyond twin control
      & \fn{unverified} & 250 \\
    \bottomrule
  \end{tabular}
\end{table}

\subsection{Cost and compatibility}

\para{Per-call latency (E3).}
\Cref{fig:eval}(a) in \cref{sec:eval-overhead} shows the per-call latency behind \cref{tab:overhead}: bars are the median of the per-pass P50s over three passes of 200 calls after 40 warm-up calls, and whiskers reach the largest P99 of the three passes.

\para{Scaling (E5).}
\Cref{fig:eval}(b,c) shows how latency grows with the number of copies per call and how memory grows with the number of resident germlines, about 12\,MB PSS per additional germline (\cref{sec:eval-overhead}).

\para{Unmodified MCP servers (E7).}
Every call to the two tools of \fn{mcp-server-time} and to \fn{mcp-server-fetch} (200 per tool) and to three tools each of \fn{mcp-server-sqlite} (\fn{list\_tables}, \fn{describe\_table}, \fn{read\_query}) and \fn{mcp-server-git} (\fn{git\_log}, \fn{git\_show}, \fn{git\_diff\_unstaged}; 30 per tool) succeeded under \sys{} with the same output as the unprotected server, 780 calls in total, apart from the clock value reported by \fn{get\_current\_time}.
\fn{mcp-server-git} ran with \fn{/dev/null} read-write in the germline; with the strict read-only binding the server cannot start, because GitPython runs \fn{git version} at import and git opens \fn{/dev/null} read-write.
The fetch tool's destination host comes from its URL argument, so the harness declared the host named there (\fn{api.example}) for each call, as \asm{5} permits.
On the cloud container, with each copy in its own PID namespace, P50 latency rises from 1.4 to 49.7\,ms for \fn{convert\_time} (15.6\,ms with a pool) and from 1.2 to 47.0\,ms for \fn{get\_current\_time} (11.2\,ms with a pool), from 18.4 to 110.3\,ms for \fn{fetch} (73.3\,ms with a pool), from 1.6--2.0 to 39.0--41.2\,ms for the sqlite tools and from 9.6--17.6 to 57.8--74.5\,ms for the git tools.

\subsection{Dampening analysis}\label{app:dampening}

\para{Rule.}
\sys{} keeps one penalty $P$ per release, which starts at zero, so an authenticated update also starts at zero.
On every call to the release, including calls refused while it is suppressed, the penalty first decays, $P\leftarrow\damp P$ with $\damp=2^{-1/\hl}$, and then grows by the weight of the call's event, if any: $w=250$ for an unverified event and 2000 for a verified witness from a release that declared determinism or for a broken determinism declaration (\cref{tab:replay}).
The release is suppressed once $P\ge S=2000$ and usable again once $P<R=750$.
The operating point is $\hl=50$ calls and $w=250$.

\para{Stationary mean.}
Let a release produce an unverified event on each call independently with probability $p$.
The expected penalty then evolves as $\mathbb{E}[P_{n+1}]=\damp\,\mathbb{E}[P_n]+wp$ and converges to the stationary mean
\begin{equation*}
  \bar P=\frac{wp}{1-\damp}\approx\frac{wp\hl}{\ln 2},
\end{equation*}
since $1-\damp\approx\ln 2/\hl$ for large $\hl$.
The mean reaches $S$ at the rate $p^{*}=S(1-\damp)/w$, which is $11.0\%$ at the operating point.
This is a statement about the mean only: the penalty of a single release fluctuates around $\bar P$, so a release whose rate lies below $p^{*}$ can still cross $S$ during a run of events, and finite-horizon probabilities must be measured, which \cref{tab:montecarlo} does.

\para{Time to suppress and recovery.}
From $P=0$, $k$ consecutive events give $P=w(1-\damp^{k})/(1-\damp)$, which first reaches $S$ at $k=9$, so a release that diverges on every call is suppressed after 9 calls.
Along the expected-penalty trajectory, in which each call adds $pw$, a release that diverges on 50\% of its calls reaches $S$ after 18 calls and one at 20\% after 58, while one at 10\% never does, because $10\%<p^{*}$.
A verified witness from a release that declared determinism adds 2000 and suppresses the release at once.
Without further events, the penalty falls from $S$ to $R$ in $\lceil\hl\log_2(S/R)\rceil=71$ calls; because refused calls also advance the decay, a suppressed release becomes usable again after 71 further call attempts, while an idle release keeps its penalty. Because the adversary controls updates and may steer the agent, it can reset the penalty with an update or speed recovery with refused calls; dampening therefore carries no guarantee, while verified witnesses remain valid evidence against the release hash they name across updates.

\para{Finite-horizon simulation.}
\Cref{tab:montecarlo} estimates the probability that a release is suppressed at least once within 1,000, 10,000 or 100,000 calls, with 2,000 simulated releases per cell (500 for 100,000 calls).
Under a random schedule, each call diverges independently with probability $p$; under a bursty schedule with the same long-run rate, divergences arrive in bursts of 10 calls, during which each call diverges with probability 0.5.
At 1\% random divergence no release is suppressed even within 100,000 calls, and at 3\% the probability is 0.15\% within 1,000 calls, 2.4\% within 10,000 and 23.2\% within 100,000.
Bursty releases are suppressed far more often at the same mean rate, because the rate inside a burst, 50\%, exceeds $p^{*}$; this is a property of the policy, which reacts to the local divergence rate rather than to the long-run average.

\begin{table}[tbp]
  \centering
  \caption{Probability (\%) that dampening suppresses a release at least once within $N$ calls, at the operating point ($\hl=50$, $w=250$, $S=2000$, $R=750$; $p^{*}=11.0\%$). Random: each call diverges independently with probability $p$. Bursty: same long-run rate, bursts of 10 calls at 50\%. ``--'': not simulated.}
  \label{tab:montecarlo}
  \footnotesize
  \setlength{\tabcolsep}{4pt}
  \begin{tabular}{@{}lrrrrrr@{}}
    \toprule
    & \multicolumn{3}{c}{Random} & \multicolumn{3}{c}{Bursty} \\
    \cmidrule(lr){2-4}\cmidrule(l){5-7}
    $p$ & $10^3$ & $10^4$ & $10^5$ & $10^3$ & $10^4$ & $10^5$ \\
    \midrule
    1\% & 0    & 0     & 0     & 11.55 & 74.35 & 100 \\
    2\% & 0.05 & 0.05  & --    & 34.95 & 99.6  & --  \\
    3\% & 0.15 & 2.4   & 23.2  & 61.25 & 100   & 100 \\
    4\% & 2.25 & 28.35 & --    & 82.2  & 100   & --  \\
    5\% & 15.8 & 86.4  & 100   & 91.85 & 100   & 100 \\
    \bottomrule
  \end{tabular}
\end{table}

\begin{figure}[tbp]
  \centering
  \safegraphic[width=\columnwidth]{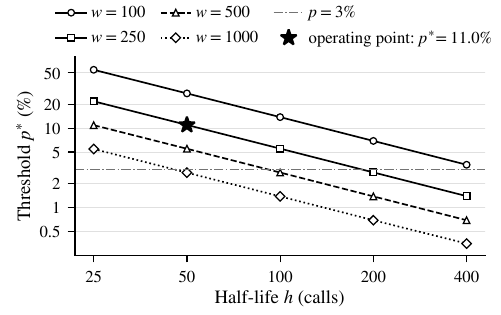}
  \caption{Dampening threshold $p^{*}$, the divergence rate at which the stationary mean penalty reaches $S$ (log scale), against half-life $\hl$ for weights $w\in\{100,250,500,1000\}$. The star marks the operating point ($\hl=50$, $w=250$, $p^{*}=11.0\%$). A line below the dash-dotted 3\% reference means the mean penalty of a release diverging on 3\% of its calls exceeds $S$; finite-horizon probabilities are in \cref{tab:montecarlo}.}
  \label{fig:dampening}
\end{figure}

\para{Parameter sweep.}
\Cref{fig:dampening} plots $p^{*}$ for half-lives of 25--400 calls and weights of 100--1000.
A line below the 3\% reference means that the stationary mean penalty of a release diverging on 3\% of its calls exceeds $S$, which happens exactly in the cells with $\hl\cdot w\ge50{,}000$; at the operating point, $\hl\cdot w=12{,}500$.
A separate simulation with 800 releases per cell shows the trade-off around the operating point ($w=250$, $S=2000$; probability of false suppression within 10,000 calls at 1\% and 3\% random divergence, and calls to suppress a release that always diverges): $\hl=25$ gives 0\% and 0\% and 9 calls ($p^{*}=22\%$); $\hl=50$ gives 0\% and 2.9\% and 9 calls ($p^{*}=11\%$); $\hl=100$ gives 0\% and 98\% and 9 calls ($p^{*}=5.5\%$). The 2.9\% at $\hl=50$ and the 2.4\% of \cref{tab:montecarlo} estimate the same probability from 800 and 2,000 releases.
Halving the weight instead ($w=125$, $\hl=100$) also gives 0\% and 0\% but needs 17 calls.
The operating point $\hl=50$ thus suppresses an always-diverging release as fast as any setting with $w=250$ while keeping false suppression at 3\% random divergence low within 10,000 calls; the same simulation at $\hl=100$ is kept in \fn{e12\_montecarlo\_h100.json}.

\FloatBarrier
\section{Ethics}\label{app:ethics}

All experiments ran in an isolated testbed on machines we control, with test data and destination hosts set up
for this study.
No third-party systems, services or users were involved.
MCPGuard-Dynamic was built from its public repository and run only on our own machines.
The disclosure we report is a threat-model boundary of destination-based confinement, which binds authority to
destinations rather than to individual items (\cref{sec:eval-mcpguard}); it does not exploit an implementation
flaw in MCPGuard.
Because the finding concerns a design boundary rather than a vulnerability, no vendor notification was needed.
The signed attack releases and the generated-variant corpus can be used to build tools that disclose data, so we
share them only through restricted access to researchers on request (\cref{app:openscience}).

\section{Artifact Availability}\label{app:openscience}

All artifacts needed to evaluate the contributions of this paper will be released with the conference version
of this paper.
The repository contains the prototype, the testbed, the experiment harnesses and the frozen result files, so that
every reported number, table and figure can be traced to a file and regenerated.

\para{Prototype.}
The artifact contains the source of \sys{}.
Its enforcement path consists of the broker and germline fork server, the per-copy confinement (namespaces,
Landlock ruleset, PID namespace per copy), \Plan{} with declared-destination splitting and the clearance split,
per-horizon state slots with their labels, and the per-copy deadline; the broker does not follow redirects.
Its evidence path consists of the twin shim, the per-destination sequence comparison with positional reply
replay, witness recording with the probe-replay \fn{verify()}, and the dampening module.
The baseline runtimes \rtSig{}, \rtIFC{}, \rtConf{}, \rtFresh{} and \rtNoSplit{} are implemented in the same code
base and selected by configuration, as are the switches that turn each mechanism off for the implementation
correspondence tests of \cref{tab:obligations,tab:regression}.
The property tests for \obl{1}--\obl{5} and the regression tests for slot labels, sequence comparison,
the per-copy deadline and redirects are included with their expected outputs.

\para{Testbed.}
The testbed consists of the 59 signed releases (the five MCP servers with their five flow-compliant releases and
52 signed updates carrying hidden code, and the two benchmark releases of \cref{sec:eval-overhead}), the test publisher key and
TUF-style metadata, the generator for the test world with its labels, clearances and per-trial canary items, the
HTTP sinks, and the disclosure oracle.
The attack variants are those of \cref{sec:threat-vectors}.

\para{Harnesses and results.}
The artifact contains the scripts that run E1--E13 and the frozen result files of every experiment in \cref{tab:eid-map}: E1, E2,
E3 (three passes), E4, E5, E6 (three repetitions), E7, E8-A, E8-C, E9 (existence bit and deniability), E10, E11,
E12, E13 (three trials), the twin test, the obligation tests on both platforms and the regression tests.
It also contains the compatibility harness for unmodified MCP servers and the two workflow tests of
\cref{sec:eval-real}, and the Monte-Carlo and parameter-sweep scripts behind \cref{app:dampening}.
\Cref{tab:eid-map} lists, for every experiment, its result file, its platform and the code version it measured.
The plotting scripts and figure sources regenerate every figure from these files.

\para{README.}
A top-level README gives a minimal runnable example, one scripted session of one tool server under a baseline
runtime and under germline execution, with its expected output, so that a reviewer can check the installation
before running the full harnesses.
It states the exact versions of the kernel, Python, the Python packages and the model-serving stack used for each
platform, the machine requirements of each harness and its run time, and a map from every table and figure of the
paper to the result file and script it is computed from.

\para{MCPGuard comparison.}
Instead of redistributing MCPGuard-Dynamic, the artifact names its public repository and the commit we built (commit \fn{825fdc0a}, 6~Oct~2026), together with
our build steps: \fn{vmlinux.h} regenerated for the server kernel, one line that registers our document server
in the proxy's server table, and one build flag that includes \fn{vmlinux.h} first; MCPGuard's enforcement code
is unchanged.
It contains our least-privilege policies in MCPGuard's format for the six-call session, which we used with
MCPGuard's proxy and its AgentBound reproduction.
It also contains the E8-A and E8-C runner scripts, the tags of the four eBPF programs as loaded in
every E8 run (\fn{file\_guard\_open}, \fn{net\_guard\_connect}, \fn{proc\_guard\_exec} and
\fn{fork\_guard\_sched\_fork}), and the positive-control transcript.
With these, a reviewer can rerun the session and observe both the blocked raw-socket transfer and the disclosure
through the authorised OCR destination host.

\para{CaMeL check.}
The artifact includes the script and its log for the labelling check of \cref{sec:eval-labels}, which runs CaMeL's reference
interpreter and Workspace policy at commit \fn{f083b6b} without a language model.

\para{Generated-variant corpus and external components.}
The E13 corpus of generated variants is our own: we generated it with a locally served model (Qwen3.5-27B) through
an orchestration pipeline built with an LLM coding agent (Codex), and we validated, signed and benchmarked every
variant ourselves (\cref{sec:eval-scale}).
The artifact contains the generation pipeline with its settings and per-attempt seeds, the E13 benchmark runner,
the validation procedure and the per-case results, so the funnel from 189 generation attempts to 91 validated
variants and the prevention rates can be checked against the result file.
Model weights (Qwen3-32B-AWQ~\cite{qwen3} and the generator) and the serving stack (vLLM~0.11~\cite{vllm}) are
obtained from their public sources and are not redistributed.

\para{Requirements.}
The prototype and E1--E5, E7 and E9--E13 run on stock Linux without kernel changes; they need PID, network and IPC
namespaces, Landlock and Python.
The cloud container (2~vCPU) ran Linux~6.18 with Landlock ABI~7 and Python~3.13; the server ran Linux~6.8 with
ABI~4 and Python~3.12.
The MCPGuard comparison (E8) additionally needs the BPF LSM enabled through the kernel boot parameter.
E6 needs GPUs to serve its model, and so does generating the E13 corpus anew with the generator model; we served the
E6 model on NVIDIA RTX 5000 Ada GPUs with 32\,GB each in a server virtual machine with 24 cores and 96\,GB RAM.
Running the E13 benchmark on the released corpus needs no GPU.
On the server, E6 takes about 40 minutes for three repetitions, one E3 pass about 1 to 3 minutes, E13 about
1 minute per trial, and E5 and E8-C under a minute each; the README lists the run time of every harness.

\para{Availability and licences.}
The package will be archived on Zenodo with a DOI.
Code is released under the MIT licence and data and results under CC~BY~4.0, with a \fn{CITATION.cff} file giving
the preferred citation.
The signed attack releases and the generated-variant corpus are released under the same licences as a
restricted-access Zenodo record, available to researchers on request (\cref{app:ethics}); all other components are
public.

\end{document}